\documentclass[preprint,12pt]{elsarticle}

\usepackage[font=small,labelfont=bf,labelsep=space]{caption}
\usepackage{amssymb}
\usepackage{amsmath,amssymb,amsthm}

\biboptions{sort&compress}

\usepackage[T1]{fontenc}
\usepackage[utf8x]{inputenc}

\usepackage[margin=2cm]{geometry}
\usepackage{mathtools}

\usepackage{upgreek}
\usepackage{subfig}

\journal{Advanced Powder Technology}

\begin{document}

\begin{frontmatter}



\title{Effect of Drag Force Modeling on the Flow of Electrostatically Charged Particles}


\author[label1,label2]{Gizem Ozler}
\author[label3]{Mustafa Demircioglu}
\author[label1,label2]{Holger Grosshans}

\affiliation[label1]{organization={Physikalisch- Technische Bundesanstalt (PTB)},
            city={Braunschweig},
            country={Germany}}
\affiliation[label2]{organization={Otto von Guericke University of Magdeburg},
            addressline={Institute of Aparatus- and Environmental Technology}, 
            city={Magdeburg},
            country={Germany}}
\affiliation[label3]{organization={Ege University},
            addressline={Institute of Chemical Engineering}, 
            city={Izmir},
            country={Turkey}}
        
\begin{abstract} 
In CFD simulations of two-phase flows, accurate drag force modeling is essential for predicting particle dynamics.
However, a generally valid formulation is lacking, as all available drag force correlations have been established for specific flow situations.
In particular, these correlations have not been evaluated for particle-laden flows subjected to electrostatic forces.
The paper reports the effect of drag force modeling on the flow of electrically charged particles.
To this end, we implemented different drag force correlations to the open-source CFD tool pafiX.
Then, we performed highly-resolved Direct Numerical Simulations (DNS) using the Eulerian-Lagrangian approach of a particle-laden channel flow with the friction Reynolds number of 180. 
The simulations generally revealed a strong influence of the precise drag
correlation on particles in the near-wall region and a minor effect on the particles far from the walls.
Due to their turbophoretic drift, particles accumulate close to the channel walls.
For uncharged particles, the simulations show large deviations of the particle concentration profile in the near-wall region depending on the drag force correlation.
Therefore, the disturbance of the flow surrounding a particle by a nearby wall or other particles is important for its drag.
Driven by electrostatic forces, charged particles accumulate even closer to the wall.
Contrary to the uncharged cases, when the particles carry a high charge (in our case one femto-coulomb), we found minor effects of drag force modeling on particle concentration profiles.
In conclusion, for the investigated conditions, we propose to account for the effect of nearby particles and walls on the drag of low- or uncharged particles.
\end{abstract}



\begin{keyword}

Particle-laden flow \sep DNS \sep  Drag Force \sep Triboelectric Charging

\end{keyword}

\end{frontmatter}



\section{Introduction}
\label{intro}
\renewcommand*{\today}{September 15, 2022}  
Particle-laden flows, consisting of particulate and a carrier fluid phase, appear in various industrial processes such as pneumatic conveying of solids, energy conversion \cite{mallouppas2013large}, or fluidized bed reactors \cite{tabaeikazerooni2019laminar}.
During these processes, particles acquire electrostatic charge through particle-particle and particle-wall collisions.
The electrification of the particles leads to the formation of deposits, dangerous sparks, and dust explosions that cause serious property damage or loss of life.
On the other hand, electrostatic charging is used in electrostatic precipitators \cite{HIDY2003273} and powder coating \cite{Prasad}. 
Therefore, knowledge of the motion of charged particles can help mitigate hazards and improve particular industrial devices.
Experimental studies of charged particle-laden flows are difficult.
It seems impossible to assign a defined charge to each particle.
Contrary, due to uncertainties in the initial and boundary conditions and the electrification mechanism \cite{grosshans2017direct}, \cite{matsusaka2010triboelectric}, the charge carried by particles is unknown.
Further, the flow region very close to a wall, which is most important for particle electrification, is challenging to visualize.
Therefore, in recent years, the focus has shifted to developing numerical tools that can easily overcome these uncertainties.

To simulate the electrification of particles during pneumatic conveying, computational fluid dynamics (CFD) tools employing the Eulerian-Lagrangian approach are the most common \cite{grosshans2017direct}, \cite{LI2021103542}. 

Therein, the Eulerian framework is used to solve the Navier-Stokes equations for the fluid phase.
The Lagrangian framework is used for computing the trajectory of each particle.
Particles are considered as a point mass, which means that particles are smaller than the flow length scale, and the boundary layer flow around particles is not resolved.
Instead, fluid velocities are interpolated from the Eulerian grid nodes to the particle locations.
The momentum transfer from the particles to the fluid is considered by using source terms.
The major interaction force between the two phases, required to obtain particle trajectories, is the aerodynamic drag.
However, since the flow around the particle is not resolved in the Eulerian-Lagrangian approach, the drag force cannot be computed directly by integrating the forces exerted on the particles' surface by the fluid.
Instead, the drag force is accounted for by appropriate closure models.
Thus, accurate drag force models are a prime measure to reduce the overall error of the simulation.

Often, the estimation of the drag force is based on the classical drag force correlation by Putnam \cite{putnamintegratable}, which is strictly valid for uniform flow. 
However, there are many effects (e.g., other particles, nearby walls) that cause non-uniformity and need to be considered to correct the drag force modeling, and thus, the simulation results. 
The goal of this work is to apply an accurate drag force correlation that can capture the mentioned physical effects, specifically for the flow of electrostatically charged particles.
We selected drag force correlations and tested those models with our CFD tool pafiX~\cite{grosshans2017direct}.
The paper is structured as follows; Section 2 reviews the literature on drag force correlations.
Section 3 outlines the mathematical model, simulation setup, and algorithm of our CFD solver.
Finally, in Section 4, simulation results are given and discussed.

\section{Literature review of drag force models}

Interphase interactions take place in many chemical engineering processes.
Modeling these interactions is essential for developing more accurate simulation tools.
In most particle-laden flow systems, the drag force is the major force acting on a particle.
Therefore, the precision of the drag force models has a direct impact on the accuracy of simulations \cite{lapple1940calculation}.

Drag force arises from contact between a fluid and an immersed body when there is a relative velocity between them.
Drag force can be separated into two types; skin friction drag (also called viscous drag), and pressure drag (also called form drag).
Skin friction drag results from viscous stresses on the body that are tangential to the surface of the object.
Pressure drag occurs if the pressure stresses, which are perpendicular to the surface, are non-uniform around the body.
Integrating the pressure and viscous stresses over the surface of the body in the direction of flow gives the total drag force on a particle \cite{chorin1990mathematical},
\begin{subequations}
\begin{equation}
\text {d}F_{\text {D}}=-P\text {d}A\text {cos}\theta+\tau_\text{w}
\text {d}A\text {sin}\theta ,
\label{eq1}
\end{equation}
\begin{equation}
F_{\text {D}}= \int_{}^{} \text {d}F_{\text {D}}=\int_{}^{} 
(-P\text {d}A\text {cos}\theta+\tau_\text{w}
\text {d}A\text {sin}\theta).
\label{eq2}
\end{equation}
\end{subequations}
In this equation, the pressure and shear forces acting on a differential area, ${\text {d}A}$, are ${P\text {d}A}$ and $\tau_{\text{w}}{\text {d}A}$, and $\theta$ is the angle enclosing the outer normal of ${\text {d}A}$ in the positive flow direction.
Since the pressure and viscous stresses on an object are not available in the point-particle approach, the drag force is computed from the dynamic pressure and the drag force coefficient,$C_{\text {D}}$,
\begin{equation}
F_{\text {D}}= C_{\text {D}}\frac{\rho_\text{f}}{2}u_{\text{rel}}^{2}A ,
\label{eq3}
\end{equation}
where $\rho_{\text{f}}$ is the fluid's density and $u_{\text{rel}}$ the relative velocity between the particle and the fluid.
The object's projected area normal to flow, $A$, is for a spherical particle the cross-sectional area normal to flow, $\pi d_{\text p}^2 /4$, $d_{\text p}$ being the particle diameter.
The drag force coefficient captures the effect of the particle shape and flow conditions.
Thus, $C_\mathrm{D}$ is a function of the particle Reynolds number, $Re_\text p = u_\text {rel} d_\text p / \nu$, where $\nu$ $({\text m^\text 2}/ s)$ is the fluid's kinematic viscosity.


To determine $C_\mathrm{D}$, \citet{stokes1850effect} gave the first analytical solution to the Navier-Stokes equation at very low Reynolds numbers ($Re_\text p  \leqslant 1$), also known as creeping or Stokes flow.
In this flow regime, inertial forces are small compared to viscous forces.
Neglecting the non-linear inertial terms, the solution to the Navier-Stokes equations leads to a drag coefficient for a single spherical particle of
\begin{equation}
C_{\text {D}}=\frac{24}{Re_\text p} .
\label{eq6}
\end{equation}
Considering inertial effects for slightly higher Reynolds numbers, \citet{oseen1910uber} improved the expression to
\begin{equation}
C_{\text {D}}=\frac{24}{Re_\text p} \left(1+\frac{3}{16}Re_\text p \right).
\label{eq7}
\end{equation}
Several other analytical solution are limited to low Reynold numbers~ \cite{goldstein1929steady,proudman1957expansions,liao2002analytic}.

A great number of experimental works have been conducted to obtain the relation between higher Reynolds numbers and the drag coefficient.
Collecting extensive experimental data, \citet{lapple1940calculation} proposed the classical standard drag curve.

For a comprehensive overview of empirical drag force correlations, the Reader is referred to the review of \citet{goossens2019review}.
Most CFD tools use the drag force correlation by \citet{putnamintegratable},
\begin{equation}
C_{\text {D}} = \left\{ \begin{array}{cl}
\frac{24}{Re_\text p} \left( 1 + \frac{1}{6} {Re_\text p}^{2/3} \right) &  \ {Re_\text p} \leq 1000 \\
0.424 &  \ {Re_\text p} > 1000 \, ,
\end{array} \right. 
\label{eq8}
\end{equation}
or by \citet{Schiller1933UberDG},
\begin{equation}
{C_\text {D}} = \left\{ \begin{array}{cl}
\frac{24}{Re_\text p} \left(1+0.15{Re_\text p}^{0.687} \right) &  \ Re_\text p \leq 1000 \\
0.44 &  \ Re_\text p > 1000 \, ,
\end{array} \right. 
\label{eqSchiller}
\end{equation}
which both approximate the standard drag curve by numerical expressions.
Nevertheless, all these correlations assume the particle to be isolated.
The following sections present drag force correlations for non-isolated particles.

\subsection{Drag force modeling considering the effect of a near-wall}

A close-by wall changes the flow pattern and wake structure of a particle.
A wall slows down particles by changing the transport properties of their boundary layers \cite{zeng2009forces}.
\citet{faxen1922widerstand} achieved the first asymptotic solution for the total hydrodynamic force on a spherical particle in rectilinear motion parallel to a wall.
However, that solution is valid only for Stokes flows and for wall distances greater than radius of the particle.
\citet{goldman1967slow} and \citet{goldman1967slow2} provided an analytical solution for the case in which the gap between the particle and the wall is small compared to the radius of particle for wall-bounded flows.
To investigate wall-slip and wall-shear interaction separately, they studies two limiting cases: 
I) spherical particles moving parallel to a plane wall in a quiescent fluid II) stationary particle in a shear flow.
They solved these problems analytically and obtained modified drag coefficients.
Their work showed that the effect of the wall is important if the gap between the particle and the wall is smaller than the particle diameter and can be neglected for distances of the order of ten particle diameters.

To check the effect of a nearby-wall, we implemented the drag force correlation of \citet{zeng2009forces} into pafiX.
Using results from a large number of simulations, \citet{zeng2009forces} extended the particle Reynolds number range of Goldman’s drag force correlations.
They showed that the effect of the wall is still considerable for moderate Reynolds numbers and offered drag force closures for a particle translating parallel to the wall in a stagnant fluid and for a stationary particle in a wall-bounded linear shear flow.
The composite drag law for a stationary particle in a wall-bounded linear shear flow is given by
\begin{subequations}
\begin{equation}
C_{\text {Ds}}=C_{\text {Ds0}}(1+\alpha_{\text {s}}Re_\text p^{\beta_{\text s}}) ,
\label{eq9}
\end{equation}
with
\begin{equation}
\alpha_{\text {s}}=0.15-0.046(1-0.16\delta^2)\exp(-0.7\delta) ,
\label{eq10}
\end{equation}
\begin{equation}
\beta_{\text {s}}=0.687+0.066(1-0.76\delta^2)\exp(-\delta^{0.9}) ,
\label{eq11}
\end{equation}
\begin{equation}
C_{\text {Ds0}}=\frac{24}{Re_\text p}\left[1+0.138 \exp(-2\delta)+\frac{9}{16(1+2\delta)}\right],
\label{eq12}
\end{equation}
\end{subequations}
and $(\delta = L/d{_\text p} − 0.5)$ is the normalized gap between the particle and wall.

The corresponding drag law for a particle moving parallel to a wall in a quiescent ambient flow reads
\begin{subequations}
\begin{equation}
C_{\text {Dt}}=C_{\text {Dt0}}(1+\alpha_{\text {t}}Re_\text p^{\beta_{\text t}}) ,
\label{eq13}
\end{equation}
where;
\begin{equation}
\alpha_{\text {s}}=0.15(1-\exp(-\sqrt{\delta}) ,
\label{eq14}
\end{equation}
\begin{equation}
\beta_{\text {t}}=0.687+0.313 \exp(-2\sqrt{\delta}) ,
\label{eq15}
\end{equation}
\begin{equation}
C_{\text {Dt0}}=\frac{24}{Re_\text p}\left[ 1.02+\frac{0.07}{1+4\delta^2}
-\frac{8}{5} \ln\left(\frac{270\delta}{135+256\delta}\right) \right].
\label{eq16}
\end{equation}
\end{subequations}
The correlations are valid for $\delta \to 0$ and converge to the correlation of \citet{Schiller1933UberDG} in the limit $\delta \to \infty $.

\subsection{Drag force modeling for non-isolated particles}
\label{sec:2.2}

Also, the presence of surrounding particles alters the fluid flow and changes the drag force on particles~\cite{ergun1949fluid,wen1966mechanics}.
Thus, in the case of dense flows, the drag force correlation needs to take into account other particles (called crowding or swarming effect).
Many experimental and numerical studies offered drag force correlations considering these effects, often using different definitions of the Reynolds number and drag force.
To avoid confusion, we define both before discussing correlations.

In a system of a steady-state fluid flow and a cloud of $N_\text {p}$ particles, the fluid exerts two forces on the particles; 
drag force, $\textbf{\textit{F}}_\text D$, (due to solid-fluid friction) and buoyancy force, $\textbf{\textit{F}}_\text B$, (due to static pressure gradient).
The sum of these two forces, the total force on the particles, $\textbf{\textit{F}}_\text T$, and the overall pressure gradient of the fluid, $\textbf{$\nabla $}P$, is related by \cite{van2005lattice}
\begin{subequations}
\begin{equation}
-\nabla P=\frac{N_\text {p}}{V_\text {sys}} \textbf{\textit{F}}_\text T =
\frac{N_\text {p}}{V_\text {sys}} (\textbf{\textit{F}}_\text D+\textbf{\textit{F}}_\text B) = \frac{N_\text {p}}{V_\text {sys}} (\textbf{\textit{F}}_\text D - V_\text {p}\nabla P) ,
\label{eq17}
\end{equation}
\begin{equation}
\textbf{\textit{F}}_\text T = \frac{1}{1-\phi}\textbf{\textit{F}}_\text D .
\label{eq18}
\end{equation}
\end{subequations}

Therein, $V_\text {sys}$ is the volume of the system, $V_\text {P}$ the volume of a single particle, and $\phi$ the solid volume fraction.

Further, the drag force in this section is normalized by Stokes drag force, i.e.,
\begin{equation}
\left|F_\text {D}  \right|=\frac{\textbf{\textit{F}}_\text {D}}{3\pi \mu d_\text {p} \textbf{\textit{U}}}
\label{eq19}
\end{equation}
where $\textbf{\textit{U}}=\varepsilon\left| \textbf{\textit{u}}_\text {p}-\textbf{\textit{u}}\right|$ is the superficial velocity with $\varepsilon$ being the void fraction, $\textbf{\textit{u}}_\text {p}$ the particle velocity and  $\textbf{\textit{u}}$ is the fluid velocity. 

Drag force correlations for non-isolated cases are mostly based on one of two different sets of equations.
The first one combines the correlations of \citet{kozeny1927uber} and \citet{cornell1953flow} to
\begin{equation}
{F_\text {D}}(\phi,Re_\text p)={F_\text {D}}(\phi,0)+{f}(\phi,Re_\text p).
\label{eq20}
\end{equation}
The term $F_\text {D}(\phi,0)$ represents the drag force in the Stokes flow limit, the term ${f}(\phi,Re_\text p)$ the effect of inertial forces.
The second one is based on the closure of \citet{dallavalle1948micromeritics},
\begin{equation}
{F_\text {D}}(\phi,Re_\text p)={F_\text {D}}(0,Re_\text p)(1-\phi)^{-\beta} \, ,
\label{eq21}
\end{equation}
where ${F_\text {D}}(0,Re_\text p)$ accounts for the particle Reynolds number of isolated particles, and $\beta$ for the crowding effect.

\citet{ergun1949fluid} proposed one of the earliest semi-empirical drag force correlations considering other particles derived from pressure drop experiments in fixed beds with zero particle velocity.
The equation of \citet{ergun1949fluid} is based on \citet{kozeny1927uber} and \citet{carman1956flow}, and is valid only for dense systems $(\phi \gg  0.20)$ of low or intermediate Reynolds numbers.
The normalized form of Ergun’s drag force reads
\begin{equation}
{F_\text {D}}=\frac{150(1-\varepsilon)}{18\varepsilon^2}+1.75\frac{Re_\text p}{18\varepsilon^2}.
\label{eq22}
\end{equation}
According to \citet{ergun1949fluid}, the total energy loss (pressure drop) is the sum of the kinetic energy loses due to inertial forces and viscous energy losses.
The first and the second terms in the equation represent the viscous and Reynolds dependent inertial forces.

\citet{wen1966mechanics}, using settling experiments, proposed a drag force correlation based on the closure of \citet{dallavalle1948micromeritics},
\begin{equation}
{F_\text {D}}=\frac{Re_\text p}{24}{C_\text {D}}\varepsilon^{-3.65} .
\label{eq23}
\end{equation}
They computed $C_\mathrm{D}$ based on Eq.~(\ref{eqSchiller}). Their correlation is valid for dilute systems equals the relation of \citet{Schiller1933UberDG} in the limit $\varepsilon\to 1$.

Combining the correlations of \citet{wen1966mechanics} and \citet{ergun1949fluid}, \citet{gidaspow1994multiphase} obtain a hybrid drag force correlation valid for both dilute and dense flows,
\begin{equation}
{F_\text {D}}(\varepsilon, Re_\text p) = \left\{ \begin{array}{cl}
\frac{150(1-\varepsilon)}{18\varepsilon^2}+1.75\frac{Re_\text p}{18\varepsilon^2} &  \ \varepsilon \leq 0.8 \\
\frac{Re_\text p}{24}{C_\text {D}}\varepsilon^{-3.65} &  \ \varepsilon > 0.8
\end{array} \right. .
\label{eq24}
\end{equation}
Again, $C_\mathrm{D}$ is calculated by Eq.~(\ref{eqSchiller}).
The correlation covers the entire range of void fractions, but contains a discontinuity at $\varepsilon=0.8$.
To ensure continuity \citet{gobin2003fluid} proposed the combination
\begin{equation}
{F_\text {D}}= \left\{ \begin{array}{cl}
{F_\text {D(Wen, Yu)}} &  \ \phi \leq 0.3 \\
\min({F_\text {D(Wen, Yu)}},{F_\text {D(Ergun, Orning)}}) &  \ \phi > 0.3 \, .
\end{array} \right.
\label{eq25}
\end{equation}

\citet{di1994voidage} examined how the drag force deviates from the drag force in the presence of other particles in an experiment in which a fluid flows through a packed fluidized bed with homogeneously distributed particles.
This let to a closure in which $\beta$ is based on $Re_\mathrm{p}$, namely
\begin{equation}
{F_\text {D}}=\frac{Re_\text p}{24}{C_\text {D}}\varepsilon^{-\beta},
\label{eq23}
\end{equation}
\begin{equation}
{\beta}=3.7-0.65 \exp\left[  -\frac{(1.5-\log(Re_\text p))^2}{2}\right].
\label{eq23}
\end{equation}

With the increase in computational power, simulations using the Lattice Boltzmann Method (LBM) or the Immersed Boundary Method (IBM) to provide closure models for complex systems became feasible.
LBM simulations were used by \citet{hill2001moderate} for monodisperse static particle packings of simple cubic, face-centered cubic, and random arrays of spheres.
Their correlation covers solid volume fractions of $0.001 \le \phi \le  0.6$ for low Reynolds numbers.
They later extended their work to moderate Reynolds numbers, up to $Re_\text p\ll 100$ \cite{hill2001first}.
However, their functions are not continuous and lack high Reynolds numbers.

\citet{benyahia2006extension} covered a broad range of Reynolds numbers and volume fractions, by modifying the correlation of \citet{hill2001moderate}.
Also, they ensured the continuity of the correlation.
\citet{van2005lattice} proposed an expression using LBM for both mono- and bi-disperse sphere packings and low Reynolds numbers $(Re_\text p\ll 1)$.
Their work was extended by \citet{beetstra2007drag} to higher Reynolds numbers $(Re_\text p\ll 1000)$ for monodispersed particles, resulting in
\begin{equation}
\begin{split}
{F_\text {D}}(\phi,Re_\text p)=\frac{10\phi}{(1-\phi)^2}+(1-\phi)^2(1+1.5\phi^{0.5})+ \\
& \hspace{-4cm}\frac{0.413Re_\text p}{24(1-\phi^2)}
\left[\frac{(1-\phi^{-1}+3\phi(1-\phi)+8.4Re_\text p^{-0.343}}{{1+10^{3\phi}}Re_\text p^{-0.5+2\phi}}  \right] \, .
\label{eq31}
\end{split}
\end{equation} 

Finally, correlations appeared that suit the conditions of pneumatic conveying, namely dilute flows of a low particle Reynolds number.
In the following, the correlations that we have implemented into pafiX and tested are given.

\citet{Tang}, using IBM, simulated the flow past fixed assemblies of monodisperse spheres in a face-centered-cubic array and for random distributions.
Their correlations build upon the closures of \citet{kozeny1927uber} and \citet{carman1956flow} and cover solid volume fractions from zero to 0.6 and particle Reynolds numbers between~50 and~1000,
\begin{equation}
\begin{split}
{F_\text {D}}(\phi,Re_\text p)=\frac{10\phi}{(1-\phi)^2}+(1-\phi)^2(1+1.5\phi^{0.5})+ \\
& \hspace{-5cm} \left[ 0.11\phi(1+\phi)-\frac{0.00456}{(1-\phi)^4}+0.169(1-\phi)+\frac{0.0644}{(1-\phi)^4}{Re_\text p}^{-0.343} \right]Re_\text p \, .
\label{eq32}
\end{split}
\end{equation}
\citet{kravets2019new}, using LBM, provided a drag force correlation for static random sphere packings.
The correlation is applicable to solid volume fractions between zero and 0.4 and particle Reynolds numbers up to 500,
\begin{equation}
\begin{split}
{F_\text {D}}(\phi,Re_\text p)=\frac{10\phi}{(1-\phi)^2}+(1-\phi)^2(1+1.5\phi^{0.5})+ \\
& \hspace{-6cm} \left[ 0.1695\phi(1+\phi)-\frac{0.004321}{(1-\phi)^4}+0.0719(1-\phi)+\frac{0.02169}{(1-\phi)^4}{Re_\text p}^{-0.2017} \right]Re_\text p.
\label{eq33}
\end{split}
\end{equation}

\section{Mathematical model and simulation setup}

\subsection{Mathematical Model}

\begin{figure}[b]
\centering
\includegraphics[width=10cm]{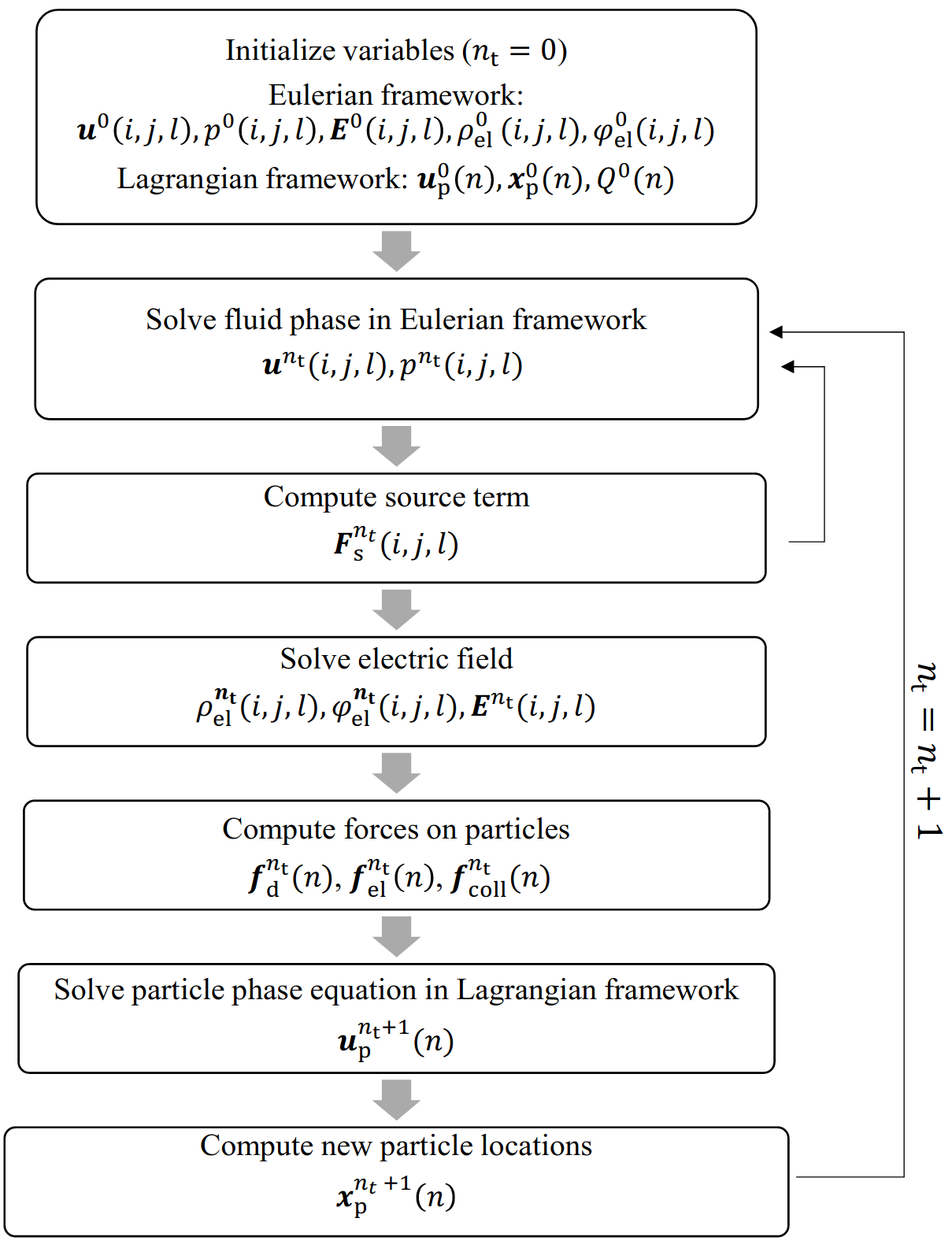}
\caption{Algorithm of CFD solver.}

\label{fig:algorithm}
\end{figure}

For this study, we use the open-source CFD tool pafiX.
The structure of the CFD solver is sketched in Fig.~\ref{fig:algorithm} and outlined in the following.

We simulated a turbulent particle-laden flow consisting of a Newtonian carrier fluid with $N_\mathrm{p}$ particles using a four-way coupled Eulerian-Lagrangian approach.
The fluid phase is governed by Navier-Stokes equations and solved in Eulerian framework.
The flow is assumed to be incompressible, so the equations for the fluid phase read
\begin{equation}
\mathbf{\nabla} \cdot\textbf{\textit{u}}=0 ,
\label{eq3.2}
\end{equation}
\begin{equation}
\frac{\partial {\textbf{\textit{u}}}}{\partial t}+
(\textbf{\textit{u}} \cdot \mathbf{\nabla})\textbf{\textit{u}}=-\frac{1}{\rho_\text f}{\nabla}P+\nu\nabla ^{2}{\textbf{\textit{u}}}+\textbf{\textit{F}}_\text s .
\label{eq3.2}
\end{equation}
Here, $\textbf{\textit{u}}$ is the fluid's velocity vector, $\rho_\text f$ its density, $P$ its pressure, and $\nu$ its kinematic viscosity.
The two phases are four-way coupled.
In four-way coupling, the interaction between the particles and the fluid and between the particles is taken into account.
The momentum transfer from the particles to the fluid is considered by the source term $\textbf{\textit{F}}_\text s$.

All particles in our simulations are rigid, spherical, and of equal material density.
The point-mass approach is employed, which means the particles are smaller than the grid cells and the effect of particles on fluid flow is taken into account by source terms.
Each particle is tracked individually in the Lagrangian framework.
The acceleration of a single particle is calculated based on the specific forces acting on it,
\begin{equation}
\frac{\partial {\textbf{\textit{u}}}_\text p}{\partial t}=\textbf{\textit{f}}_{\text{ad}}+\textbf{\textit{f}}_{\text{el}}+\textbf{\textit{f}}_{\text{coll}}+\textbf{\textit{f}}_{\text{g}},
\label{eq3.2}
\end{equation}
where $\textbf{\textit{u}}_\text p$ is the particle velocity, $\textbf{\textit{f}}_{\text {ad}}$ the aerodynamic drag, $\textbf{\textit{f}}_{\text {el}}$ the particle acceleration due to the electric field, $\textbf{\textit{f}}_{\text {coll}}$ the collisional acceleration, and $\textbf{\textit{f}}_{\text {g}}$ the acceleration due to the net effect of gravity.
For the aerodynamic acceleration,
\begin{equation}
\textbf{\textit{f}}_{\text {ad}}=-\frac{3\rho_\text f}{8\rho_\text p r_\text p}C_\text d\left| \textbf{\textit{u}}_\text {rel} \right|\textbf{\textit{u}}_\text {rel},
\label{eq3.2}
\end{equation}
the default version of pafiX uses the drag coefficient of \citet{putnamintegratable}.
The acceleration due to electric field  $\textbf{\textit{f}}_{\text {el}}$ is calculated as
\begin{equation}
\textbf{\textit{f}}_{el} = \frac{Q\textbf{\textit{E}}}{\text m_{\text p}},
\label{eq8}
\end{equation}
\begin{equation}
\textbf{\textit{E}} = \mathbf{\nabla}\varphi, 
\label{Elect}
\end{equation}
where $Q$ is the particle's charge, $\textbf{\textit{E}}$ the electrical field strength and $\varphi$ the electrical potential which satisfies the Poisson equation $\nabla ^{2}\varphi=\rho_{\text {el}}/\epsilon$, where $\rho_\text {el}$ is the electrical charge density and $\epsilon$ the permittivity of the fluid. Eqn.~\ref{Elect} and the Poisson equation are discretized and solved numerically via a second-order central difference scheme.
The permittivity of fluid is assumed to equal the permittivity of vacuum.

For a system with control volume of $V$, the sum of the charges of $N$ number of particles is given as
\begin{equation}
\int_{}^{}\rho_{\text{el}}\text{d}V=\sum_{i=1}^{N}Q_{i}. 
\label{eq8}
\end{equation}

For the fluid phase, second-order central difference schemes approximate the convective, viscous, and pressure gradient terms.
The particle trajectories are integrated by a second-order Crank-Nicolson scheme.

\subsection{Simulation setup}

\begin{figure}[h]
\centering
\includegraphics[width=12cm]{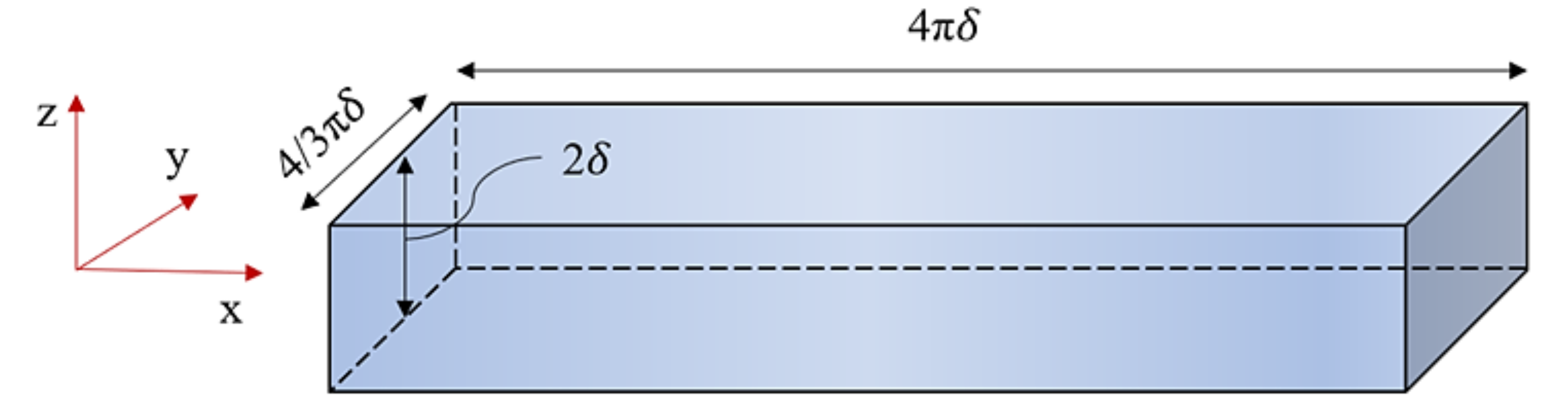}
\caption{Schematical representation of the computational domain.}
\label{fig:domain}
\end{figure}

We simulated electrically charged and uncharged particle-laden flows in a channel between two parallel planar walls.
A constant external pressure gradient drove a fluid flow of a friction Reynolds number $Re_{\tau} = 180$.
The friction Reynolds number is defined as $Re_{\tau} = \frac{u_{\tau} \delta}{\nu}$ with $\delta$ being the channel half-width, $u_{\tau}$ friction velocity ($u_{\tau} =\sqrt{\tau_{w}/\rho_{\text f}}$), and $\tau_{\text w}$ the shear stress at the wall.
Uniformly distributed particles were seeded inside the fully developed turbulent flow. 
The computational domain is sketched in Fig.~\ref{fig:domain}.

In our computational domain, the $x-$, $y-$, and $z-$axes points in the streamwise, spanwise and wall-normal directions.
The dimensions of the computational domain are $4\pi \delta$, $4/3\pi \delta$, and $2\delta$.
Periodic boundary conditions are prescribed in streamwise and spanwise directions.

Based on a grid resolution study by \citet{grosshans2021effect} for the same flow conditions, we chose a grid of 256$\times$144$\times$144 cells in $x-$, $y-$, and $z-$directions.
The cell's size is uniform in $x-$ and $y-$directions and refined towards the walls in $z-$direction.

We set the particle properties consistent to the simulations of \citet{sardina2012wall}.
Accordingly, the material density of the particles is 924~$\text {kg}/\text m^3$.
The particles' Stokes number, which is defined as
\begin{equation}
St^{+}=\frac{\tau_{\text{p}}u_{\tau}^{2}}{\nu} \, ,
\label{eq8}
\end{equation}
and the particle response time $\tau_{\text{p}}$ as
\begin{equation}
\tau_{\text{p}}=\frac{\rho_{\text{p}}d_{p}^{2}}{18\rho_{\text{f}}\nu} \,
\label{eq8}
\end{equation}
is set to $St^{+} = 1$.
Thus, the particles are monodisperse and of a diameter of $16.9~\upmu$m.
Inelastic collisions are considered by a restitution coefficient of 0.9.

We assumed that particles do not exchange electrostatic charge through particle-particle or particle-wall collisions;
instead, the and particle charge is constant.
Also, to investigate the isolated influence of the drag force, we neglected gravitational and lift forces.

\section{Results and discussion}

In the following, we discuss the simulation results for uncharged and charged particles.
We show the evolving particle concentrations and the effect of charge on them.
Then, we present the results using different drag force correlations for uncharged and charged particles and elaborate on the interplay between the drag and electrostatic forces.

\subsection{Uncharged particles}

\begin{figure}[t]
\centering
\includegraphics[width=9cm]{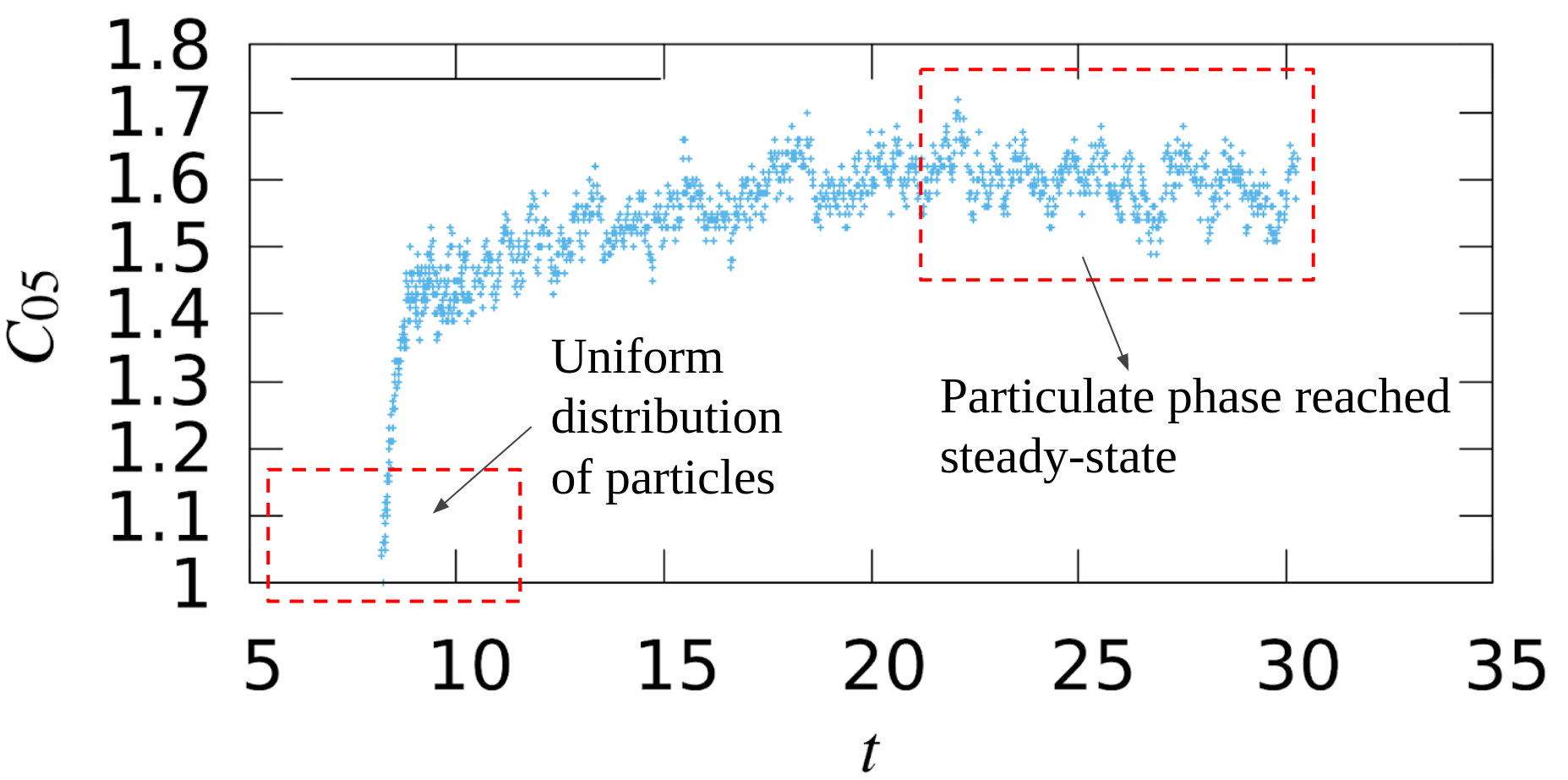}
\caption{Temporal evolution from the initially homogeneous to the steady-state concentration of uncharged particles.}
\label{fig:evolconc1}
\end{figure}

In this section, we investigate, for uncharged particles, the temporal evolution of the concentration profiles, the flow pattern after reaching steady-state, and the effect of drag force correlations.
Fig.~\ref{fig:evolconc1} shows the evolution of the particle concentrations until reaching their steady-state using the drag force correlation of \citet{putnamintegratable}.

As mentioned above, we started the simulations from a fully developed turbulent flow and seeded uniformly distributed particles.
The initial particle velocity equals that of the fluid at that location.
The figure plots the concentration of particles within the 5\% of the computational domain closest to the wall normalized to the overall particle concentration.
Since many physical phenomena occur in the near-wall region, $C_\mathrm{05}$ is a good indicator to check the simulation's temporal convergence.

Due to the initially uniform particle distribution, the initial normalized particle concentration in Fig.~\ref{fig:evolconc1} equals unity.
The plot shows the particles' tendency to accumulate in the near-wall region, even though the rate of accumulation decreases with time.

The migration of particles toward the wall is the result of "turbophoresis", which is discussed in depth below.
The particle concentration using other drag force correlations follows qualitatively the same pattern.

The simulation for the given case reached convergence at about $t=17$~s.
After particles have reached the steady-state, the results of the simulations are averaged in space and time to eliminate random turbulent fluctuations.

\begin{figure}[h]
\centering
\subfloat[]{\includegraphics[width=0.48\textwidth]{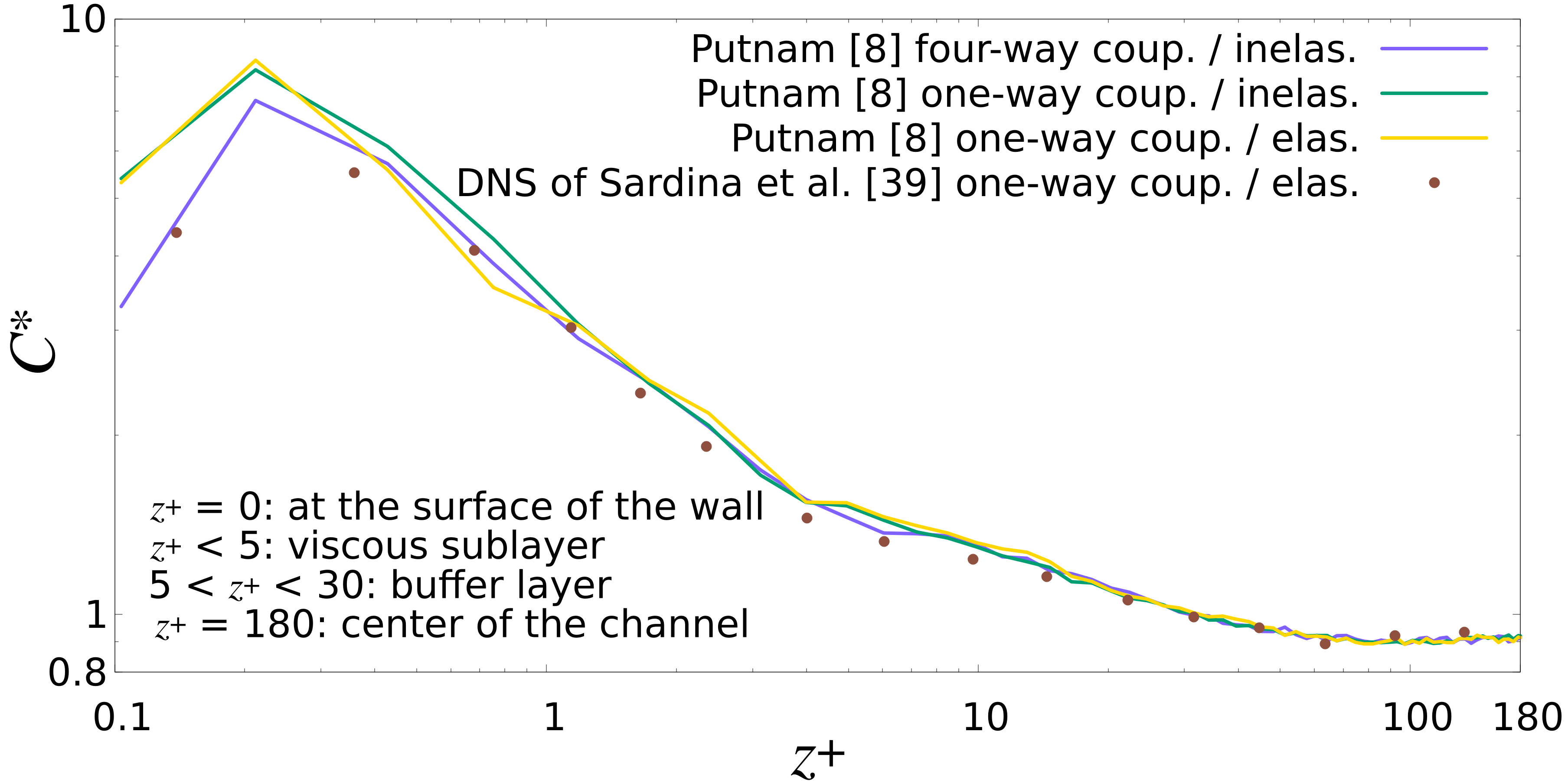}\label{fig:Sardinacoupling}}
\quad
\subfloat[]{\includegraphics[width=0.48\textwidth]{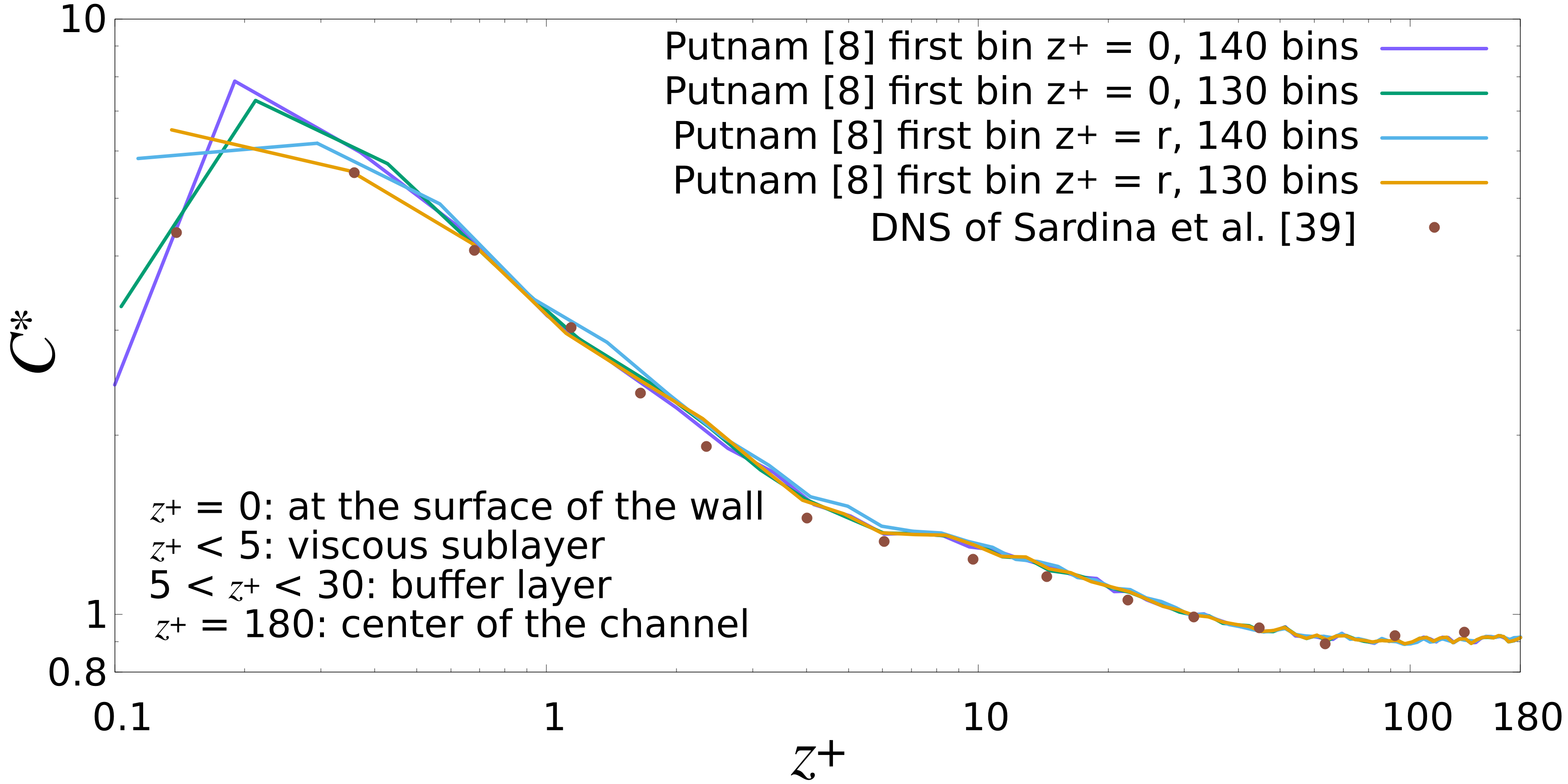}\label{fig:Sardinabins}}

\caption[]{
Comparison of our simulations with the DNS of \citet{sardina2012wall} depending on (a) particle interactions and (b) post-processing.}
\label{fig:78}
\end{figure}

To validate our simulation tool for the particulate phase, Fig.~\ref{fig:Sardinacoupling} and Fig.~\ref{fig:Sardinabins} compares our results to the DNS of \citet{sardina2012wall}.
The distance from the wall, $z^+$, is normalized by viscous length-scale, $\delta_{\nu}$.
We compare with our simulation using the drag force correlation of \citet{putnamintegratable} since that is the correlation implemented by \citet{sardina2012wall}.
In general, our simulations agree nearly completely with those of the reference paper.
Differences occur only within $z^+<1$, which is an extremely thin band close to the wall.
Nevertheless, we investigate the reason for these small differences.

Fig.~\ref{fig:Sardinacoupling} and Fig.~\ref{fig:Sardinabins} highlights the influences of discrepancies in the physical modeling of \citet{sardina2012wall} and us.
First, four-way coupling approach is used in our simulation, i.e., we model the effect of particles on fluid and vice-versa and particle-particle interactions.
Contrary, in the reference work, one-way coupling is implemented.
To understand if the differences in the coupling approach lead to the deviation in the near-wall region, we performed a simulation case with one-way coupling approach.
As can be seen in Fig.~\ref{fig:Sardinacoupling} , our simulation with the one-way coupling approach deviates even more from the simulation results of \citet{sardina2012wall}.
Inter-particle collisions tend to move particles from high to low concentration regions. 
Neglecting inter-particle collisions leads to the increase of the particle concentration in the high concentration region $y^+<1$.

Next, in our simulation the collision of particles with the wall is inelastic with a restitution coefficient of 0.9, i.e., the kinetic energy of a particle is not conserved but lost to internal friction during collisions.
Contrary, in the reference paper, purely elastic collisions are simulated.
\citet{li2001numerical} reported that the difference between restitution coefficient of 0.9 and elastic collisions does not change the simulation results.
However, taking into account lift and gravity forces, they based their conclusion on a different simulation setup than ours.

Thus, to clear up the effect of the restitution coefficient on our simulations, we simulated a case with purely elastic collisions.
As confirmed by Fig.~\ref{fig:Sardinacoupling}, also in our set-up, the elasticity of the collisions do not influence the final particle concentration profiles.
In conclusion, the small differences in the physical of \citet{sardina2012wall} and us do not cause the differences in the resulting particle concentrations in $y^+<1$.

Finally, we investigated the influence of post-processing on the resulting profiles.
More specifically, we tried different arrangements and sizes of the bins with which we generated the concentration profiles.
As visualized in Fig.~\ref{fig:Sardinabins}, the details of the bins make significant differences in the concentration profile in the near-wall region.
In all spatial profiles analyzing particles presented in this paper, we process the location of the particles' center point.
The first bin starts at $z^+ = r^+$ because the particles' center point cannot move closer to a wall.
Using this definition, the first two curves in Fig.~\ref{fig:Sardinabins} perfectly agree with the data point closest to the wall of \citet{sardina2012wall}.
This data point agrees even for a reduction of the number of bins from 140 to 130.
However, if we start the first bin directly at the wall ($z^+ = 0$), the second data point coincides with the one of \citet{sardina2012wall}.
Beyond the second data point, the curves anyhow agree.
Notably, if the first bin directly starts at the wall ($z^+ = 0$) and the number of bins reduces to 130, the concentration peak appears to attach to the wall.
However, this apparent attachment is not physical but results from flawed post-processing.

In conclusion, the concentration profiles in the vicinity of the wall ($z^+<1$) are sensitive to details of the post-processing procedure.
These details explain the discrepancies between the reference and our data.
In the remainder of this paper, profiles are generated with 140~bins starting from $z^+=r^+$, which is the arrangement that we consider to reflect the flow pattern the most accurately.

\begin{figure}[t]
\centering
\subfloat[]{\includegraphics[width=0.48\textwidth]{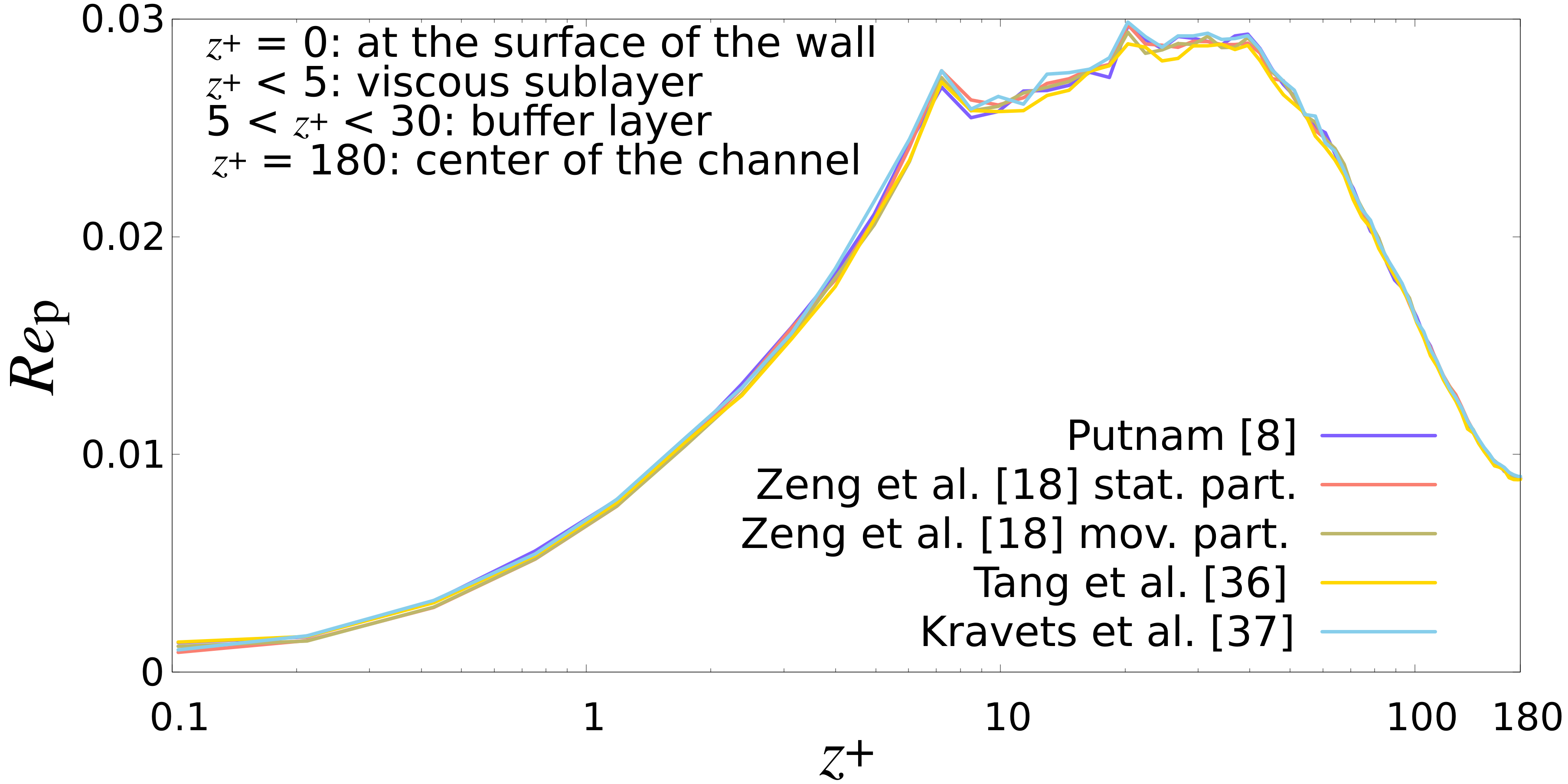}\label{fig:reunch}}
\quad
\subfloat[]{\includegraphics[width=0.48\textwidth]{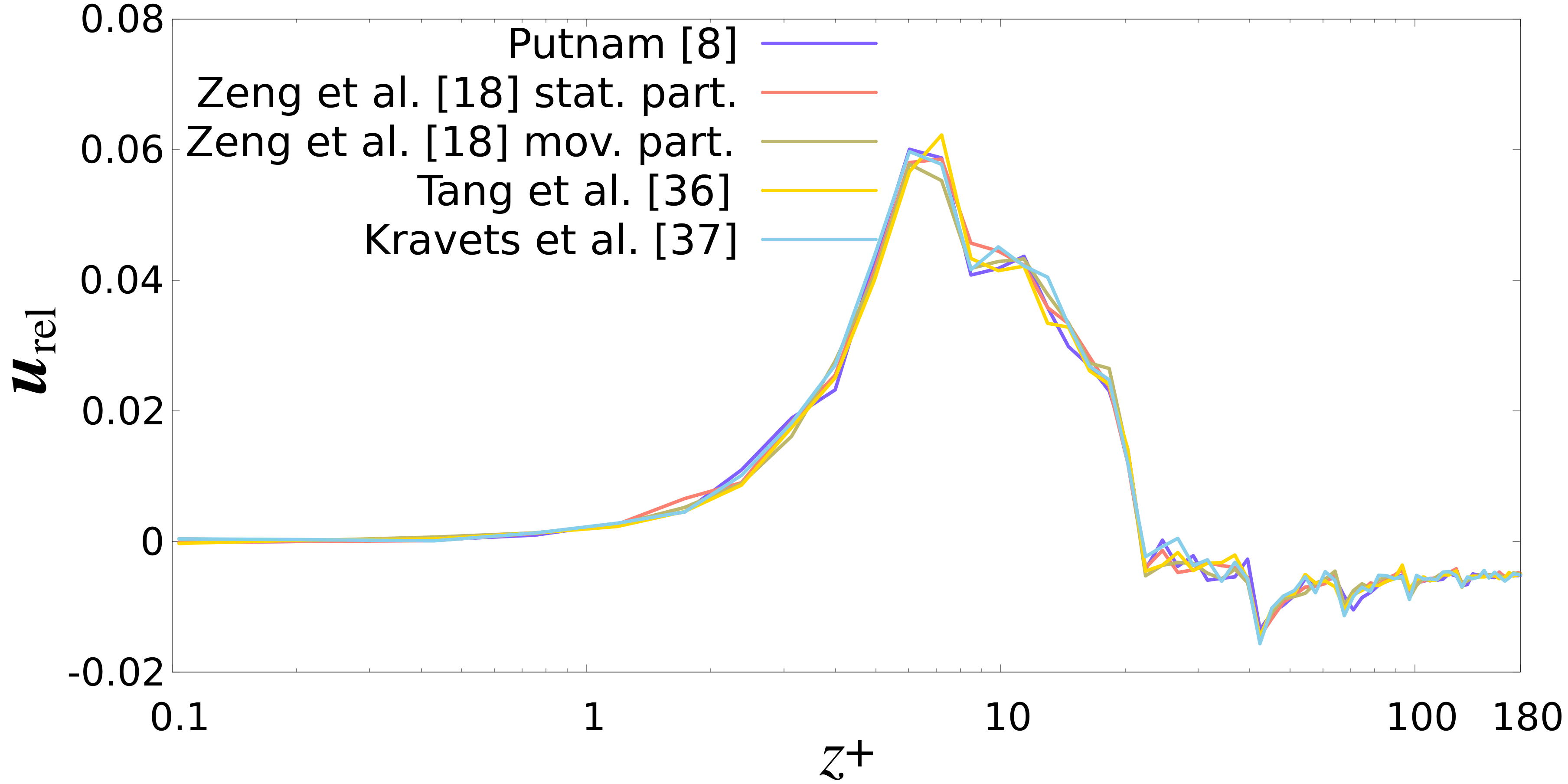}\label{fig:velunch}}
\caption{Profiles of the (a) particle Reynolds number and (b) relative velocity for different drag correlations.}
\end{figure}

Next, we explore the influence of the drag force correlations.
For all investigated drag force correlations, the average particle Reynolds number during the steady-state phase as a function of the distance from the wall is given in Fig.~\ref{fig:reunch}.
The drag force correlations for non-isolated particles are computed based on the particle Reynolds numbers using the superficial velocity, $Re_\text p = (1-\phi) u_\text {rel} d_\text p / \nu$ (see Section~\ref{sec:2.2}).
However, for better comparison, in Fig.~\ref{fig:reunch} we plot all particle Reynolds numbers using the relative velocity independent of the fluid volume fraction.

According to the plot, all drag force correlations give the same particle Reynolds number profiles.
For all simulations, $Re_\text p$ is lower than 0.1, thus, Stoke’s flow regime is valid.
The particle Reynolds number is very low in the near-wall region and increases steadily up to $z^+=7$.
It continues to increase with fluctuations until $z^+ = 13$ and, then, decreases toward the center of the channel.

Additionally, as the particle Reynolds number depends on the absolute value of the relative velocity between fluid and particle $u_{\text {rel}}$, we plot in Fig.~\ref{fig:velunch} the corresponding profiles of the relative streamwise velocity, $u_{\text {rel}}$ with respect to normalized distance from the wall.
Due to low absolute fluid and particle velocities near the wall, also their relative velocity is low;
it starts to increase in the range $1<z^+<7$.
Toward the center, the relative velocity decreases.
And beyond $z^+>12$ it is even lower than in viscous sublayer.
Therefore, a decrease in the relative velocity for $z^+ > 40$ leads to a decrease in the particle Reynolds number in that region.

\begin{figure}[b]
\centering
\includegraphics[width=10cm]{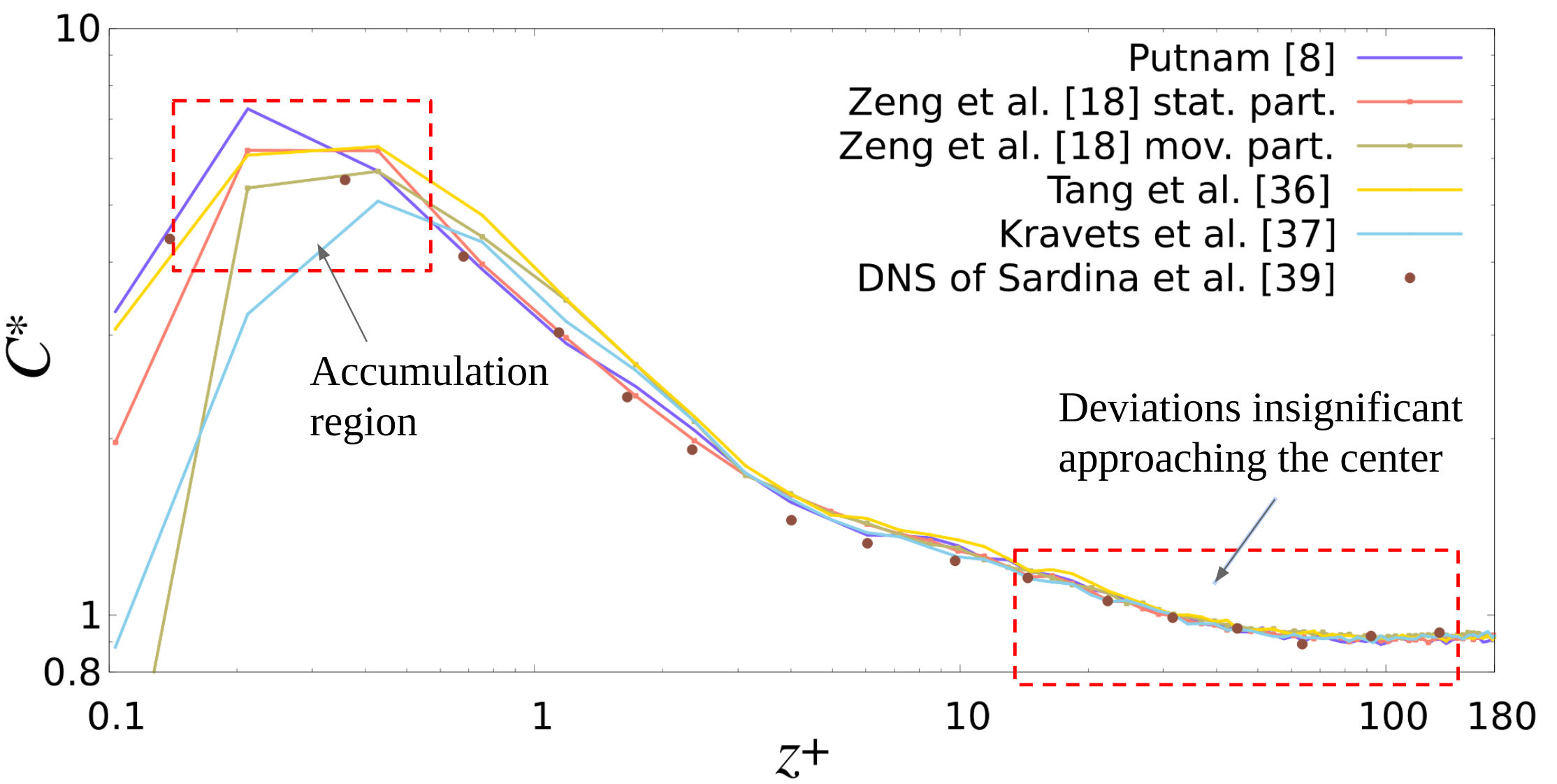}
\caption{Influence of the drag force correlation on the concentration profiles of uncharged particles.}
\label{fig:draguncharc}
\end{figure}

In Fig.~\ref{fig:draguncharc}, we compare concentration profiles using different drag force correlations.
Before discussing the differences between drag force relations, we look at the general increase in the concentration of particles close to the wall.
Particles accumulate in the near the wall region as a result of turbophoresis, the tendency of particles to travel from regions of high to regions of lower turbulent kinetic energy.
Due to the local difference in turbulent energy, in wall-bounded flows particle migration occurs, which leads to long-term accumulation in specific regions, mostly in viscous sublayer.
On the other hand, a fraction of particles that deposit in the near-wall region moves directly to the wall by diffusion from the accumulation region, which smoothes the concentration curve and creates a peak in the accumulation region \cite{marchioli2002mechanisms}.

According to Fig.~\ref{fig:draguncharc}, different correlations lead to deviating concentration profiles, especially in the near-wall region.
Concentration deviations using correlations based on surrounding particles \cite{kravets2019new,Tang} can be attributed to the higher particle concentration and non-uniformity of the particle distribution in the near-wall region.

Compared to the default correlation by \citet{putnamintegratable}, the drag correlations modeling the effect of a wall \cite{zeng2009forces} decrease the particle concentration in the near-wall region.
As Fig.~\ref{fig:draguncharc} reveals, the concentration profiles obtained with the correlations of \citet{zeng2009forces}, which were established under different premisses (stationary particle in shear flow vs. moving particle in a quiescent flow) lead to differences in the near-wall region.
According to \citet{zeng2009forces}, the wall affects the drag force on particles in two different ways.
The first is due to the additional shear that the wall creates, which alters the total force on the particle.
The second one is the breaking of the symmetry around the particle due to the local acceleration of fluid in the gap between the particle and the wall.
Asymmetry around the particle alters the pressure gradient around the particle, which eventually affects the drag force.
Both ways increase the total drag force on the particle.  

At moderate distances from the wall, the concentrations obtained using different correlations become the same.
Even though Fig.~\ref{fig:draguncharc} does not allow to assess the drag correlations, the commonalities of the profiles let us conclude that drag force modeling is important in the near-wall region, especially in the viscous sublayer, but becomes insignificant beyond $z^+ \cong 5$.

\subsection{Charged particles}

\begin{figure}[h] 
\centering
\includegraphics[width=9cm]{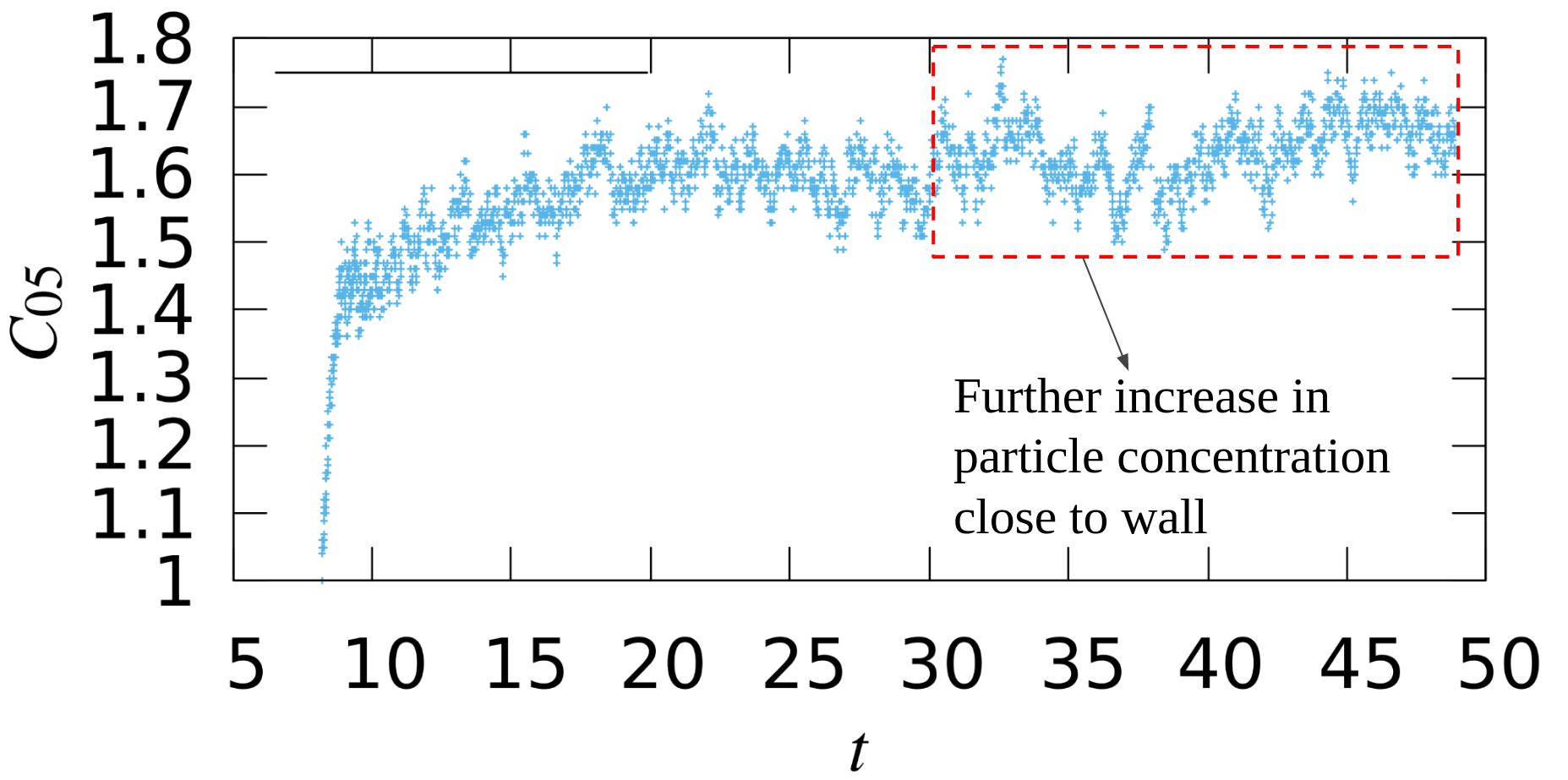}
\caption{Temporal evolution from the initially homogeneous to the steady-state concentration of charged particles.}
\label{fig:evolconc2}
\end{figure}

In this section, we present the effect of drag force correlations on the evolution of the flow pattern of charged powder.
Analogous to Fig.~\ref{fig:evolconc1}, Fig.~\ref{fig:evolconc2} plots the temporal evolution of the particle concentration close to the walls but for charged particles.
Again, the figure presents results using the drag force correlation of \citet{putnamintegratable}.

The simulation procedure, as indicated in Fig.~\ref{fig:evolconc2} was as follows:
to decrease the computational time, we restarted the simulations from the point at which the uncharged case has reached the steady-state.
Our study aims to investigate the dynamics of particles of a constant charge, isolated from charge exchange by particle-particle and particle-wall collisions.
Therefore, all particles were assigned the same charge, a constant charge.
As uncharged particles, charged particles accumulate in the near-wall region.
After assigning charge we again simulated until the particle concentrations reached steady-state.

\begin{figure}[t]
\centering
\subfloat[]{\includegraphics[width=0.48\textwidth]{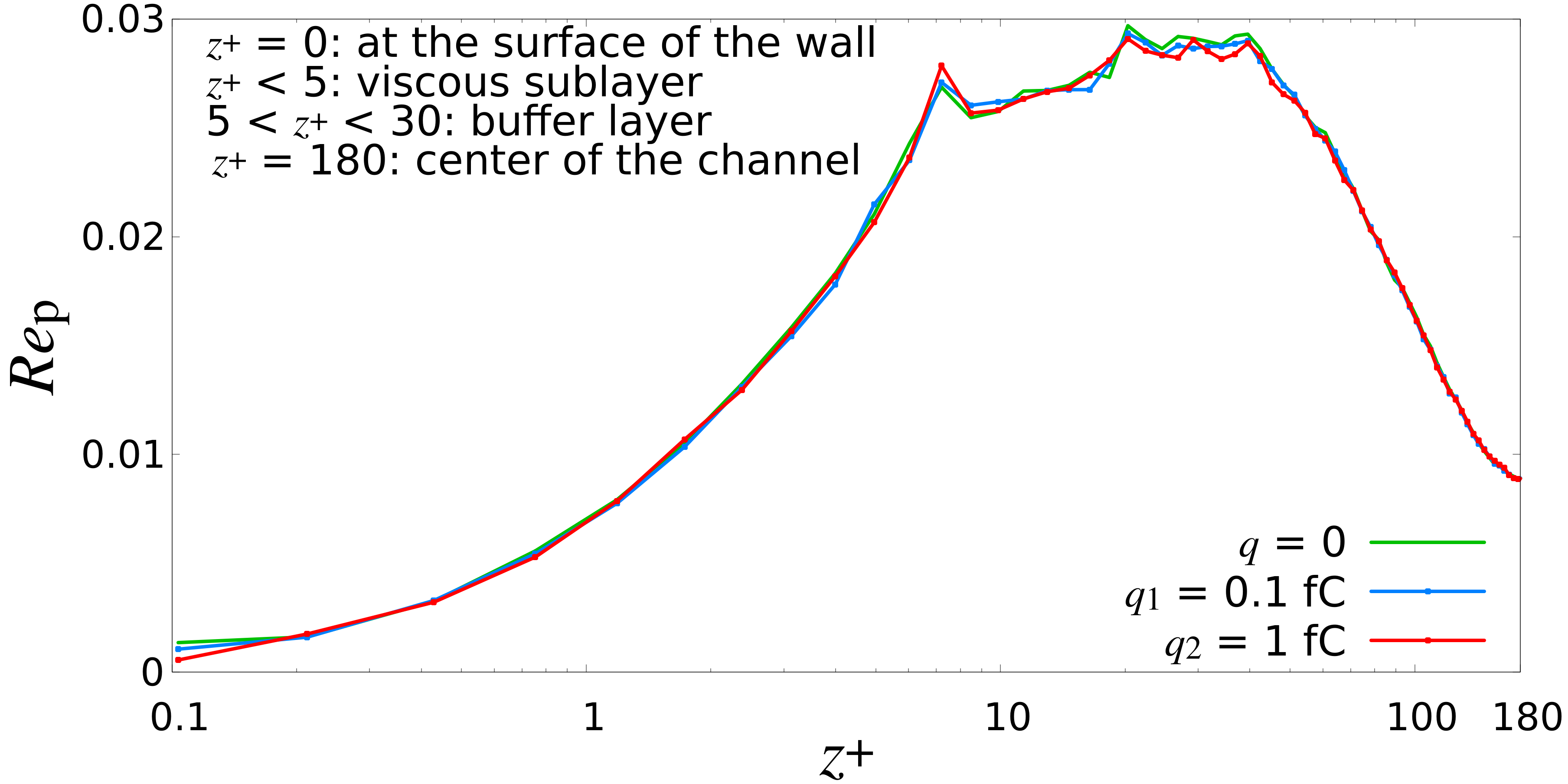}\label{fig:recharc}}
\quad
\subfloat[]{\includegraphics[width=0.48\textwidth]{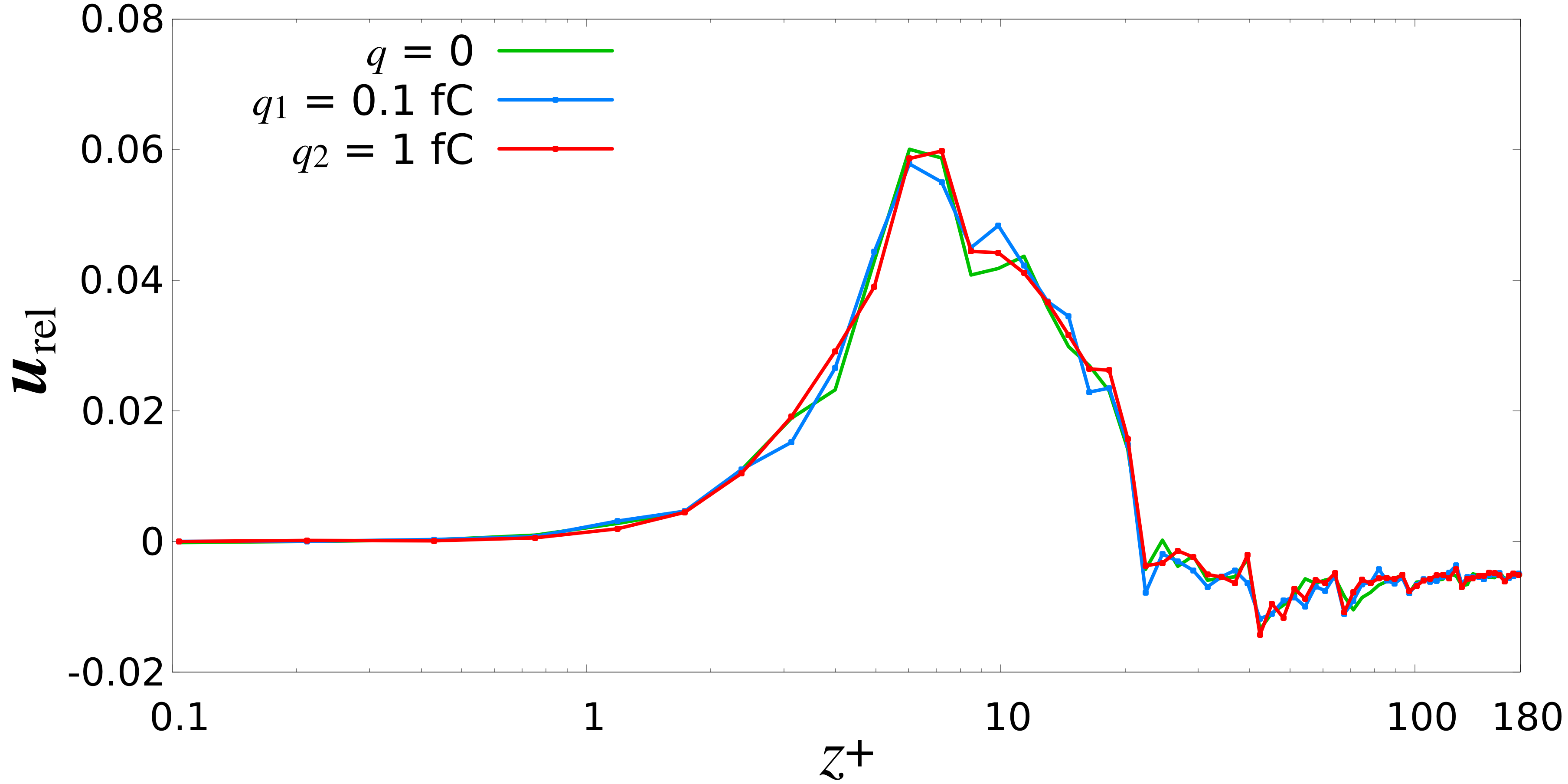}\label{fig:ucharc}}
\caption{Profiles of the (a) particle Reynolds number and (b) relative velocity for the drag correlation of \citet{putnamintegratable} and particles of different charge.}
\end{figure}

Fig.~\ref{fig:recharc} plots the particle Reynolds number with respect to normalized distance from the wall using the drag force correlation of \citet{putnamintegratable}.
The other correlations deviate only marginally;
thus, we plot only the results of one correlation to examine the general pattern.
As for the uncharged case, the particle Reynolds number is in the range of $ Re_p < 0.1$.
Also, the particle Reynolds number is low in the near-wall region and increases away from the wall.
Generally, the charge does not affect the particle Reynolds number, except very close to the wall ($z^+ \le 0.2 $) where the particle Reynolds number decreases with increasing electrostatic charge.

Independent of the charge, the particle Reynolds number increases up to $z^+ = 13$.
Close to the center, the particle Reynolds number decreases for both charged and uncharged particles as a result of decrease in relative particle-fluid velocity.
The relative velocity with respect to normalized distance from the wall is given in Fig.~\ref{fig:ucharc}.
The relative velocity profiles have the same pattern for the uncharged and charged case, only subject to statistical fluctuations.

\begin{figure}[b]
\centering
\includegraphics[width=10cm]{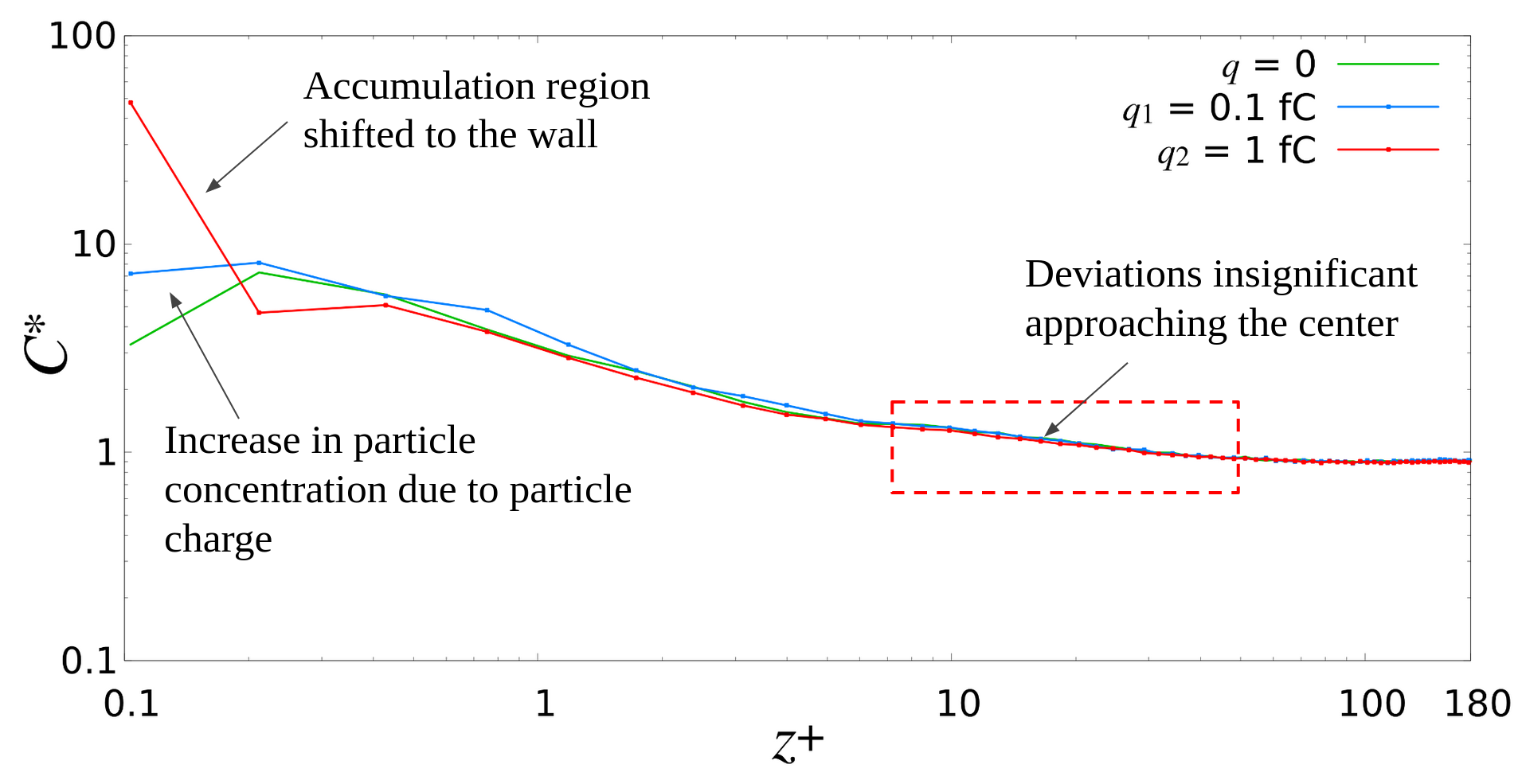}
\caption{Influence of the charge on the concentration profiles for the drag correlation of \citet{putnamintegratable}.}
\label{fig:Ccharc}
\end{figure}

In the previous section, we discussed the migration of uncharged particles toward lower turbulence energy levels due to turbophoresis, resulting in their accumulation near the wall.
To check the preferential accumulation region of charged particles, the concentration profiles for different charges are given in Fig~\ref{fig:Ccharc}.
All profiles were established using the drag correlation of \citet{putnamintegratable}. 
Both uncharged particles and particles carrying a small charge, $q_\text 1=0.1$~fC, accumulate in the viscous sublayer, their concentration peak is located at $z^+ \approx 0.2$.

Contrary, particles carrying a ten times higher charge, $q_\text 2 = 1$~fC, deposit directly at the wall.
Also, the concentration of particles in the region $0.1 < z^+ < 0.2$ was found to be higher for particles with higher charge.
For highly charged particles, electric forces dominate over turbophoretic forces and push the particles directly to the wall instead of accumulating near the wall.
Moreover, electric forces prevented particles from being transported back from the wall to the bulk flow since electric forces act in the opposite direction of turbulence diffusion.

Beyond the viscous sublayer, the particle concentration is independent of the charge.
In this region, number of particles is lower.
That means the spacing in-between particles is large and the electrical forces on each other, which reduce with the square of their distance, diminishes.
As a result of higher particle concentrations, the charge affects the particle concentration profile mostly in the near-wall region.

\begin{figure}[t]
\centering
\subfloat[$q_\text 1=0.1$~fC]{\includegraphics[width=0.48\textwidth]{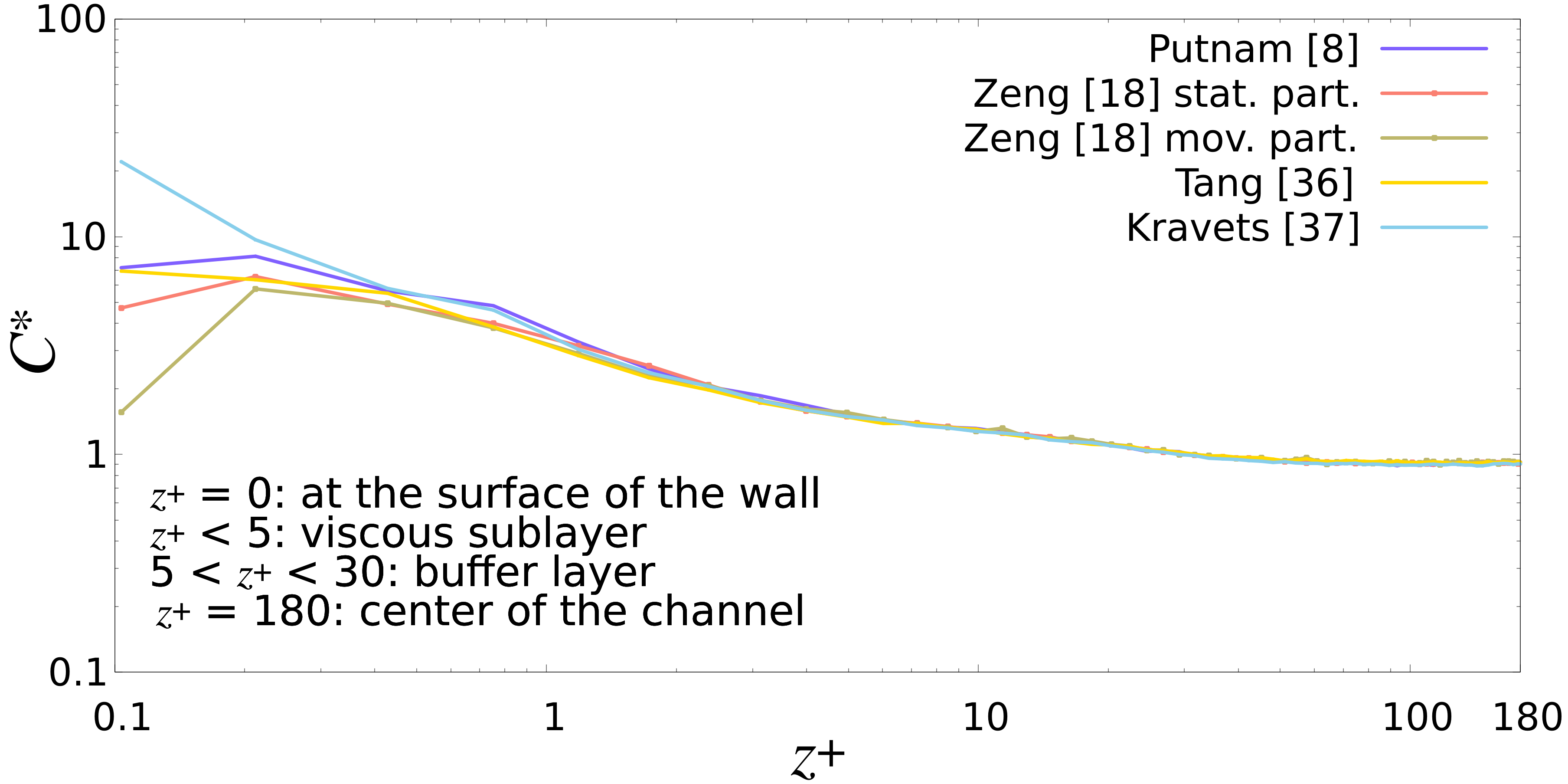}\label{fig:q1}}
\quad
\subfloat[$q_\text 2=1$~fC]{\includegraphics[width=0.48\textwidth]{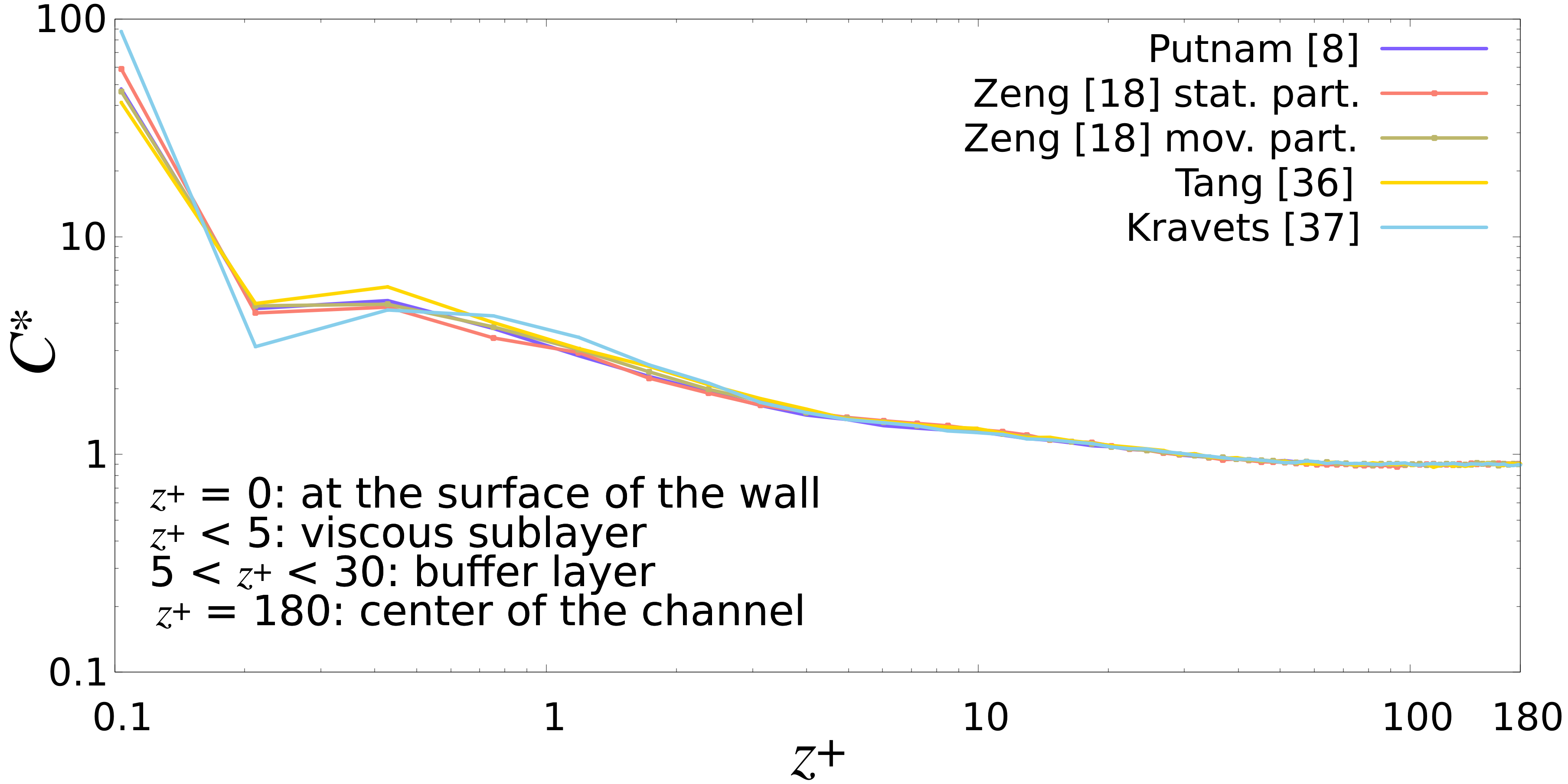}\label{fig:q2}}
\caption{Influence of the drag force correlation on the concentration profiles of charged particles.}
\end{figure}

However, the behavior observed in Fig.~\ref{fig:Ccharc} does not hold for all drag force correlations.
Fig. ~\ref{fig:q1} and Fig.~\ref{fig:q2} compared the simulation results for different drag force correlations and for two different charge levels.
For the lower charge, Fig.~\ref{fig:q1}, the particles accumulate in different regions depending on the drag force correlation.
However, for the lower charge, Fig.~\ref{fig:q2}, in all simulation the particles accumulate directly at the wall.

For the particles with the lower charge, the location of the concentration peak depends on the group of drag force correlations:
drag force correlations for particle swarms predict particle accumulation region directly at the wall.
On the other hand, drag force correlations including the near-wall effect predict the same detachment of the accumulation region for charged particles as for uncharged ones.
For charge level $q_\text 1$ not only the particle accumulation location deviates using different drag force correlations but also the concentration profile in the viscous sublayer.
Contrary, the concentration profiles for charge level $q_\text 2$ are the same independent of the drag force correlations are compatible.
Thus, with increasing particle charge the modeling of the drag force becomes less important because of the dominance of electric forces.

\section{Conclusion}

In wall-bounded flows, turbophoresis drives both charged and uncharged particles towards the wall.
In our simulations, uncharged particles accumulated in the viscous sublayer, $z^+ \approx 0.2$.
When applying a small charge of 0.1~fC, the particles still accumulate in the viscous sublayer.
But for a higher charge of 1~fC, the concentration peak attaches directly to the wall.
That means above a certain charge, electric forces dominate particle dynamics.

We investigated the effect of physical modeling on the particle concentration profiles.
One-way coupling, i.e., neglecting the momentum transfer from the particles to the fluid and other particles, showed to overestimate the particle concentration in the near-wall region compared to four-way coupling.
Besides, we found the concentration profiles to be sensitive to the post-processing procedure.
Thus, details of the post-processing procedure should be reported together with the results.

We found that the drag force models determine the numerically predicted particle concentration profiles of un- or weakly charged particles in the near-wall region, $(z^+<5)$.
Correlations that account for disturbances to the flow by nearby walls predicted the particle concentration in this region to reduce. 
However, there is no clear trend when trying different correlations considering the crowding effect.
In any case, the particle concentrations beyond the viscous sublayer are independent of the drag force model.

Contrary, for particles carrying a high charge (1~fC) the influence of the drag force model on the particle concentration profiles diminishes, even close to the wall.
Based on these results, we conclude that electrostatic forces affect particle motion by dominating the drag force when particles are sufficiently charged. Thus, the higher the particles' charge the lower the requirement for the accuracy of the drag force model.



\bibliographystyle{elsarticle-num-names} 
\bibliography{ms}

\begin{thebibliography}{41}
\expandafter\ifx\csname natexlab\endcsname\relax\def\natexlab#1{#1}\fi
\providecommand{\url}[1]{\texttt{#1}}
\providecommand{\href}[2]{#2}
\providecommand{\path}[1]{#1}
\providecommand{\DOIprefix}{doi:}
\providecommand{\ArXivprefix}{arXiv:}
\providecommand{\URLprefix}{URL: }
\providecommand{\Pubmedprefix}{pmid:}
\providecommand{\doi}[1]{\href{http://dx.doi.org/#1}{\path{#1}}}
\providecommand{\Pubmed}[1]{\href{pmid:#1}{\path{#1}}}
\providecommand{\bibinfo}[2]{#2}
\ifx\xfnm\relax \def\xfnm[#1]{\unskip,\space#1}\fi
\bibitem[{Mallouppas and van Wachem(2013)}]{mallouppas2013large}
\bibinfo{author}{G.~Mallouppas}, \bibinfo{author}{B.~van Wachem},
\newblock \bibinfo{title}{Large eddy simulations of turbulent particle-laden
  channel flow},
\newblock \bibinfo{journal}{Int. J. Multiphase Flow} \bibinfo{volume}{54}
  (\bibinfo{year}{2013}) \bibinfo{pages}{65--75}.
\bibitem[{Tabaeikazerooni(2019)}]{tabaeikazerooni2019laminar}
\bibinfo{author}{S.~H. Tabaeikazerooni}, \bibinfo{title}{Laminar and turbulent
  particle laden flows: a numerical and experimental study}, Ph.D. thesis, KTH
  Royal Institute of Technology, \bibinfo{year}{2019}.
\bibitem[{Hidy(2003)}]{HIDY2003273}
\bibinfo{author}{G.~Hidy},
\newblock \bibinfo{title}{Aerosols},
\newblock in: \bibinfo{booktitle}{Encyclopedia of Physical Science and
  Technology}, \bibinfo{edition}{third} ed., \bibinfo{publisher}{Academic
  Press}, \bibinfo{address}{New York}, \bibinfo{year}{2003}, pp.
  \bibinfo{pages}{273--299}.
  \DOIprefix\doi{https://doi.org/10.1016/B0-12-227410-5/00014-4}.
\bibitem[{Prasad et~al.(2016)Prasad, McGinity, and Robert}]{Prasad}
\bibinfo{author}{L.~Prasad}, \bibinfo{author}{J.~McGinity},
  \bibinfo{author}{W.~Robert},
\newblock \bibinfo{title}{Electrostatic powder coating: Principles and
  pharmaceutical applications},
\newblock \bibinfo{journal}{Int. J. Pharm.} \bibinfo{volume}{505}
  (\bibinfo{year}{2016}).
\bibitem[{Grosshans and Papalexandris(2017)}]{grosshans2017direct}
\bibinfo{author}{H.~Grosshans}, \bibinfo{author}{M.~V. Papalexandris},
\newblock \bibinfo{title}{Direct numerical simulation of triboelectric charging
  in particle-laden turbulent channel flows},
\newblock \bibinfo{journal}{J. Fluid Mech.} \bibinfo{volume}{818}
  (\bibinfo{year}{2017}) \bibinfo{pages}{465--491}.
\bibitem[{Matsusaka et~al.(2010)Matsusaka, Maruyama, Matsuyama, and
  Ghadiri}]{matsusaka2010triboelectric}
\bibinfo{author}{S.~Matsusaka}, \bibinfo{author}{H.~Maruyama},
  \bibinfo{author}{T.~Matsuyama}, \bibinfo{author}{M.~Ghadiri},
\newblock \bibinfo{title}{Triboelectric charging of powders: A review},
\newblock \bibinfo{journal}{Chem. Eng. Sci.} \bibinfo{volume}{65}
  (\bibinfo{year}{2010}) \bibinfo{pages}{5781--5807}.
\bibitem[{Li et~al.(2021)Li, Yao, Zhao, and Wang}]{LI2021103542}
\bibinfo{author}{J.~Li}, \bibinfo{author}{J.~Yao}, \bibinfo{author}{Y.~Zhao},
  \bibinfo{author}{C.-H. Wang},
\newblock \bibinfo{title}{Large eddy simulation of electrostatic effect on
  particle transport in particle-laden turbulent pipe flows},
\newblock \bibinfo{journal}{J. Electrostat.} \bibinfo{volume}{109}
  (\bibinfo{year}{2021}) \bibinfo{pages}{103542}.
\bibitem[{Putnam(1961)}]{putnamintegratable}
\bibinfo{author}{A.~Putnam},
\newblock \bibinfo{title}{Integratable form of droplet drag coefficient},
\newblock \bibinfo{journal}{J. Am. Rocket Soc.}  (\bibinfo{year}{1961})
  \bibinfo{pages}{1467--4798}.
\bibitem[{Lapple and Shepherd(1940)}]{lapple1940calculation}
\bibinfo{author}{C.~Lapple}, \bibinfo{author}{C.~Shepherd},
\newblock \bibinfo{title}{Calculation of particle trajectories},
\newblock \bibinfo{journal}{Ind. Eng. Chem.} \bibinfo{volume}{32}
  (\bibinfo{year}{1940}) \bibinfo{pages}{605--617}.
\bibitem[{Chorin et~al.(1990)Chorin, Marsden, and
  Marsden}]{chorin1990mathematical}
\bibinfo{author}{A.~J. Chorin}, \bibinfo{author}{J.~E. Marsden},
  \bibinfo{author}{J.~E. Marsden}, \bibinfo{title}{A mathematical introduction
  to fluid mechanics}, volume~\bibinfo{volume}{3},
  \bibinfo{publisher}{Springer}, \bibinfo{year}{1990}.
\bibitem[{Stokes(1850)}]{stokes1850effect}
\bibinfo{author}{G.~Stokes},
\newblock \bibinfo{title}{On the effect of internal friction of fluids on the
  motion of pendulums},
\newblock \bibinfo{journal}{Trans. Camb. Phil. Soc.} \bibinfo{volume}{9}
  (\bibinfo{year}{1850}) \bibinfo{pages}{106}.
\bibitem[{Oseen(1910)}]{oseen1910uber}
\bibinfo{author}{C.~W. Oseen},
\newblock \bibinfo{title}{Uber die stokes'sche {F}ormel und über eine
  verwandte {A}ufgabe in der {H}ydrodynamik},
\newblock \bibinfo{journal}{Ark. Mat. Astr. Fys.} \bibinfo{volume}{6}
  (\bibinfo{year}{1910}) \bibinfo{pages}{1}.
\bibitem[{Goldstein(1929)}]{goldstein1929steady}
\bibinfo{author}{S.~Goldstein},
\newblock \bibinfo{title}{The steady flow of viscous fluid past a fixed
  spherical obstacle at small {R}eynolds numbers},
\newblock \bibinfo{journal}{Proc. R. Soc. Lond.} \bibinfo{volume}{123}
  (\bibinfo{year}{1929}) \bibinfo{pages}{225--235}.
\bibitem[{Proudman and Pearson(1957)}]{proudman1957expansions}
\bibinfo{author}{I.~Proudman}, \bibinfo{author}{J.~Pearson},
\newblock \bibinfo{title}{Expansions at small {R}eynolds numbers for the flow
  past a sphere and a circular cylinder},
\newblock \bibinfo{journal}{J. Fluid Mech.} \bibinfo{volume}{2}
  (\bibinfo{year}{1957}) \bibinfo{pages}{237--262}.
\bibitem[{Liao(2002)}]{liao2002analytic}
\bibinfo{author}{S.-J. Liao},
\newblock \bibinfo{title}{An analytic approximation of the drag coefficient for
  the viscous flow past a sphere},
\newblock \bibinfo{journal}{Int. J. Non Linear Mech.} \bibinfo{volume}{37}
  (\bibinfo{year}{2002}) \bibinfo{pages}{1--18}.
\bibitem[{Goossens(2019)}]{goossens2019review}
\bibinfo{author}{W.~R. Goossens},
\newblock \bibinfo{title}{Review of the empirical correlations for the drag
  coefficient of rigid spheres},
\newblock \bibinfo{journal}{Powder Technol.} \bibinfo{volume}{352}
  (\bibinfo{year}{2019}) \bibinfo{pages}{350--359}.
\bibitem[{Schiller(1933)}]{Schiller1933UberDG}
\bibinfo{author}{L.~Schiller},
\newblock \bibinfo{title}{Uber die {G}rundlegenden {B}erechnungen bei der
  {S}chwerkraftaufbereitung},
\newblock \bibinfo{year}{1933}.
\bibitem[{Zeng et~al.(2009)Zeng, Najjar, Balachandar, and
  Fischer}]{zeng2009forces}
\bibinfo{author}{L.~Zeng}, \bibinfo{author}{F.~Najjar},
  \bibinfo{author}{S.~Balachandar}, \bibinfo{author}{P.~Fischer},
\newblock \bibinfo{title}{Forces on a finite-sized particle located close to a
  wall in a linear shear flow},
\newblock \bibinfo{journal}{Phys. Fluids} \bibinfo{volume}{21}
  (\bibinfo{year}{2009}) \bibinfo{pages}{033302}.
\bibitem[{Fax{\'e}n(1922)}]{faxen1922widerstand}
\bibinfo{author}{H.~Fax{\'e}n},
\newblock \bibinfo{title}{Der {W}iderstand gegen die {B}ewegung einer starren
  {K}ugel in einer z{\"a}hen {F}l{\"u}ssigkeit, die zwischen zwei parallelen
  ebenen {W}{\"a}nden eingeschlossen ist},
\newblock \bibinfo{journal}{Ann. Phys.} \bibinfo{volume}{373}
  (\bibinfo{year}{1922}) \bibinfo{pages}{89--119}.
\bibitem[{Goldman et~al.(1967{\natexlab{a}})Goldman, Cox, and
  Brenner}]{goldman1967slow}
\bibinfo{author}{A.~J. Goldman}, \bibinfo{author}{R.~G. Cox},
  \bibinfo{author}{H.~Brenner},
\newblock \bibinfo{title}{Slow viscous motion of a sphere parallel to a plane
  wall—i motion through a quiescent fluid},
\newblock \bibinfo{journal}{Chem. Eng. Sci.} \bibinfo{volume}{22}
  (\bibinfo{year}{1967}{\natexlab{a}}) \bibinfo{pages}{637--651}.
\bibitem[{Goldman et~al.(1967{\natexlab{b}})Goldman, Cox, and
  Brenner}]{goldman1967slow2}
\bibinfo{author}{A.~Goldman}, \bibinfo{author}{R.~G. Cox},
  \bibinfo{author}{H.~Brenner},
\newblock \bibinfo{title}{Slow viscous motion of a sphere parallel to a plane
  wall—ii couette flow},
\newblock \bibinfo{journal}{Chem. Eng. Sci.} \bibinfo{volume}{22}
  (\bibinfo{year}{1967}{\natexlab{b}}) \bibinfo{pages}{653--660}.
\bibitem[{Ergun and Orning(1949)}]{ergun1949fluid}
\bibinfo{author}{S.~Ergun}, \bibinfo{author}{A.~A. Orning},
\newblock \bibinfo{title}{Fluid flow through randomly packed columns and
  fluidized beds},
\newblock \bibinfo{journal}{Ind. Eng. Chem.} \bibinfo{volume}{41}
  (\bibinfo{year}{1949}) \bibinfo{pages}{1179--1184}.
\bibitem[{Wen(1966)}]{wen1966mechanics}
\bibinfo{author}{C.~Y. Wen},
\newblock \bibinfo{title}{Mechanics of fluidization},
\newblock in: \bibinfo{booktitle}{Chem. Eng. Prog. Symp. Ser.},
  volume~\bibinfo{volume}{62}, \bibinfo{year}{1966}, pp.
  \bibinfo{pages}{100--111}.
\bibitem[{van~der Hoef et~al.(2005)van~der Hoef, Beetstra, and
  Kuipers}]{van2005lattice}
\bibinfo{author}{M.~A. van~der Hoef}, \bibinfo{author}{R.~Beetstra},
  \bibinfo{author}{J.~Kuipers},
\newblock \bibinfo{title}{Lattice-{B}oltzmann simulations of
  low-{R}eynolds-number flow past mono-and bidisperse arrays of spheres:
  results for the permeability and drag force},
\newblock \bibinfo{journal}{J. Fluid Mech.} \bibinfo{volume}{528}
  (\bibinfo{year}{2005}) \bibinfo{pages}{233--254}.
\bibitem[{Kozeny(1927)}]{kozeny1927uber}
\bibinfo{author}{J.~Kozeny},
\newblock \bibinfo{title}{Uber kapillare {L}eitung der {W}asser in {B}oden},
\newblock \bibinfo{journal}{R. Acad. Sci., Vienna, Proc. Class I}
  \bibinfo{volume}{136} (\bibinfo{year}{1927}) \bibinfo{pages}{271--306}.
\bibitem[{Cornell and Katz(1953)}]{cornell1953flow}
\bibinfo{author}{D.~Cornell}, \bibinfo{author}{D.~L. Katz},
\newblock \bibinfo{title}{Flow of gases through consolidated porous media},
\newblock \bibinfo{journal}{Ind. Eng. Chem.} \bibinfo{volume}{45}
  (\bibinfo{year}{1953}) \bibinfo{pages}{2145--2152}.
\bibitem[{DallaValle(1948)}]{dallavalle1948micromeritics}
\bibinfo{author}{J.~M. DallaValle},
\newblock \bibinfo{title}{Micromeritics: the technology of fine particles}
  (\bibinfo{year}{1948}).
\bibitem[{Carman(1956)}]{carman1956flow}
\bibinfo{author}{P.~C. Carman},
\newblock \bibinfo{title}{Flow of gases through porous media}
  (\bibinfo{year}{1956}).
\bibitem[{Gidaspow(1994)}]{gidaspow1994multiphase}
\bibinfo{author}{D.~Gidaspow}, \bibinfo{title}{Multiphase flow and
  fluidization: continuum and kinetic theory descriptions},
  \bibinfo{publisher}{Academic press}, \bibinfo{year}{1994}.
\bibitem[{Gobin et~al.(2003)Gobin, Neau, Simonin, Llinas, Reiling, and
  S{\'e}lo}]{gobin2003fluid}
\bibinfo{author}{A.~Gobin}, \bibinfo{author}{H.~Neau},
  \bibinfo{author}{O.~Simonin}, \bibinfo{author}{J.-R. Llinas},
  \bibinfo{author}{V.~Reiling}, \bibinfo{author}{J.-L. S{\'e}lo},
\newblock \bibinfo{title}{Fluid dynamic numerical simulation of a gas phase
  polymerization reactor},
\newblock \bibinfo{journal}{Int. J. Numer. Methods Fluids} \bibinfo{volume}{43}
  (\bibinfo{year}{2003}) \bibinfo{pages}{1199--1220}.
\bibitem[{Di~Felice(1994)}]{di1994voidage}
\bibinfo{author}{R.~Di~Felice},
\newblock \bibinfo{title}{The voidage function for fluid-particle interaction
  systems},
\newblock \bibinfo{journal}{Int. J. Numer. Methods Fluids} \bibinfo{volume}{20}
  (\bibinfo{year}{1994}) \bibinfo{pages}{153--159}.
\bibitem[{Hill et~al.(2001{\natexlab{a}})Hill, Koch, and
  Ladd}]{hill2001moderate}
\bibinfo{author}{R.~J. Hill}, \bibinfo{author}{D.~L. Koch},
  \bibinfo{author}{A.~J. Ladd},
\newblock \bibinfo{title}{Moderate-{R}eynolds-number flows in ordered and
  random arrays of spheres},
\newblock \bibinfo{journal}{J. Fluid Mech.} \bibinfo{volume}{448}
  (\bibinfo{year}{2001}{\natexlab{a}}) \bibinfo{pages}{243--278}.
\bibitem[{Hill et~al.(2001{\natexlab{b}})Hill, Koch, and Ladd}]{hill2001first}
\bibinfo{author}{R.~J. Hill}, \bibinfo{author}{D.~L. Koch},
  \bibinfo{author}{A.~J. Ladd},
\newblock \bibinfo{title}{The first effects of fluid inertia on flows in
  ordered and random arrays of spheres},
\newblock \bibinfo{journal}{J. Fluid Mech} \bibinfo{volume}{448}
  (\bibinfo{year}{2001}{\natexlab{b}}) \bibinfo{pages}{213--241}.
\bibitem[{Benyahia et~al.(2006)Benyahia, Syamlal, and
  O'Brien}]{benyahia2006extension}
\bibinfo{author}{S.~Benyahia}, \bibinfo{author}{M.~Syamlal},
  \bibinfo{author}{T.~J. O'Brien},
\newblock \bibinfo{title}{Extension of {H}ill--{K}och--{L}add drag correlation
  over all ranges of {R}eynolds number and solids volume fraction},
\newblock \bibinfo{journal}{Powder Technol.} \bibinfo{volume}{162}
  (\bibinfo{year}{2006}) \bibinfo{pages}{166--174}.
\bibitem[{Beetstra et~al.(2007)Beetstra, van~der Hoef, and
  Kuipers}]{beetstra2007drag}
\bibinfo{author}{R.~Beetstra}, \bibinfo{author}{M.~A. van~der Hoef},
  \bibinfo{author}{J.~Kuipers},
\newblock \bibinfo{title}{Drag force of intermediate {R}eynolds number flow
  past mono-and bidisperse arrays of spheres},
\newblock \bibinfo{journal}{AICHE. J.} \bibinfo{volume}{53}
  (\bibinfo{year}{2007}) \bibinfo{pages}{489--501}.
\bibitem[{Tang et~al.(2014)Tang, Kriebitzsch, Peters, Hoef, and Kuipers}]{Tang}
\bibinfo{author}{Y.~Tang}, \bibinfo{author}{S.~Kriebitzsch},
  \bibinfo{author}{E.~Peters}, \bibinfo{author}{M.~Hoef},
  \bibinfo{author}{H.~Kuipers},
\newblock \bibinfo{title}{A new drag correlation from fully resolved
  simulations of flow past monodisperse static arrays of spheres},
\newblock \bibinfo{journal}{AICHE. J.} \bibinfo{volume}{61}
  (\bibinfo{year}{2014}). \DOIprefix\doi{10.1002/aic.14645}.
\bibitem[{Kravets et~al.(2019)Kravets, Rosemann, Reinecke, and
  Kruggel-Emden}]{kravets2019new}
\bibinfo{author}{B.~Kravets}, \bibinfo{author}{T.~Rosemann},
  \bibinfo{author}{S.~Reinecke}, \bibinfo{author}{H.~Kruggel-Emden},
\newblock \bibinfo{title}{A new drag force and heat transfer correlation
  derived from direct numerical {LBM}-simulations of flown through particle
  packings},
\newblock \bibinfo{journal}{Powder Technol.} \bibinfo{volume}{345}
  (\bibinfo{year}{2019}) \bibinfo{pages}{438--456}.
\bibitem[{Grosshans et~al.(2021)Grosshans, Bissinger, Calero, and
  Papalexandris}]{grosshans2021effect}
\bibinfo{author}{H.~Grosshans}, \bibinfo{author}{C.~Bissinger},
  \bibinfo{author}{M.~Calero}, \bibinfo{author}{M.~V. Papalexandris},
\newblock \bibinfo{title}{The effect of electrostatic charges on particle-laden
  duct flows},
\newblock \bibinfo{journal}{J. Fluid Mech.} \bibinfo{volume}{909}
  (\bibinfo{year}{2021}).
\bibitem[{Sardina et~al.(2012)Sardina, Schlatter, Brandt, Picano, and
  Casciola}]{sardina2012wall}
\bibinfo{author}{G.~Sardina}, \bibinfo{author}{P.~Schlatter},
  \bibinfo{author}{L.~Brandt}, \bibinfo{author}{F.~Picano},
  \bibinfo{author}{C.~M. Casciola},
\newblock \bibinfo{title}{Wall accumulation and spatial localization in
  particle-laden wall flows},
\newblock \bibinfo{journal}{J. Fluid Mech.} \bibinfo{volume}{699}
  (\bibinfo{year}{2012}) \bibinfo{pages}{50--78}.
\bibitem[{Li et~al.(2001)Li, McLaughlin, Kontomaris, and
  Portela}]{li2001numerical}
\bibinfo{author}{Y.~Li}, \bibinfo{author}{J.~B. McLaughlin},
  \bibinfo{author}{K.~Kontomaris}, \bibinfo{author}{L.~Portela},
\newblock \bibinfo{title}{Numerical simulation of particle-laden turbulent
  channel flow},
\newblock \bibinfo{journal}{Phys. Fluids} \bibinfo{volume}{13}
  (\bibinfo{year}{2001}) \bibinfo{pages}{2957--2967}.
\bibitem[{Marchioli and Soldati(2002)}]{marchioli2002mechanisms}
\bibinfo{author}{C.~Marchioli}, \bibinfo{author}{A.~Soldati},
\newblock \bibinfo{title}{Mechanisms for particle transfer and segregation in a
  turbulent boundary layer},
\newblock \bibinfo{journal}{J. Fluid Mech.} \bibinfo{volume}{468}
  (\bibinfo{year}{2002}) \bibinfo{pages}{283--315}.

\end{thebibliography}





\end{document}